\documentclass[twoside,twocolumn,english]{revtex4-1}

\usepackage[T1]{fontenc}
\usepackage[latin9]{inputenc}
\usepackage{geometry}
\usepackage{verbatim}
\usepackage{amsmath}
\usepackage{amssymb}
\usepackage{graphicx}
\usepackage{float}

\usepackage{color,soul}

\usepackage{enumerate}

\usepackage{mathbbol}
\usepackage{amsfonts}
\usepackage{ulem}

\makeatletter

\RequirePackage{colortbl, tabularx}
\@ifundefined{comment}{}
  {
   
  }%

\makeatother

\usepackage{babel}

\usepackage{color}

\definecolor{Green}{rgb}{0.10,0.60,0.00}

\setstcolor{red}

\begin{document}

\global\long\def\l{\lambda}%
\global\long\def\ints{\mathbb{Z}}%
\global\long\def\nat{\mathbb{N}}%
\global\long\def\re{\mathbb{R}}%
\global\long\def\com{\mathbb{C}}%
\global\long\def\dff{\triangleq}%
\global\long\def\df{\coloneqq}%
\global\long\def\del{\nabla}%
\global\long\def\cross{\times}%
\global\long\def\der#1#2{\frac{d#1}{d#2}}%
\global\long\def\bra#1{\left\langle #1\right|}%
\global\long\def\ket#1{\left|#1\right\rangle }%
\global\long\def\braket#1#2{\left\langle #1|#2\right\rangle }%
\global\long\def\ketbra#1#2{\left|#1\right\rangle \left\langle #2\right|}%
\global\long\def\paulix{\begin{pmatrix}0  &  1\\
 1  &  0 
\end{pmatrix}}%
\global\long\def\pauliy{\begin{pmatrix}0  &  -i\\
 i  &  0 
\end{pmatrix}}%
\global\long\def\pauliz{\begin{pmatrix}1  &  0\\
 0  &  -1 
\end{pmatrix}}%
\global\long\def\sinc{\mbox{sinc}}%
\global\long\def\ft{\mathcal{F}}%
\global\long\def\dg{\dagger}%
\global\long\def\bs#1{\boldsymbol{#1}}%
\global\long\def\norm#1{\left\Vert #1\right\Vert }%
\global\long\def\H{\mathcal{H}}%
\global\long\def\tens{\varotimes}%
\global\long\def\rationals{\mathbb{Q}}%
 
\global\long\def\tri{\triangle}%
\global\long\def\lap{\triangle}%
\global\long\def\e{\varepsilon}%
\global\long\def\broket#1#2#3{\bra{#1}#2\ket{#3}}%
\global\long\def\dv{\del\cdot}%
\global\long\def\eps{\epsilon}%
\global\long\def\rot{\vec{\del}\cross}%
\global\long\def\pd#1#2{\frac{\partial#1}{\partial#2}}%
\global\long\def\L{\mathcal{L}}%
\global\long\def\inf{\infty}%
\global\long\def\d{\delta}%
\global\long\def\I{\mathbb{I}}%
\global\long\def\D{\Delta}%
\global\long\def\r{\rho}%
\global\long\def\hb{\hbar}%
\global\long\def\s{\sigma}%
\global\long\def\t{\tau}%
\global\long\def\O{\Omega}%
\global\long\def\a{\alpha}%
\global\long\def\b{\beta}%
\global\long\def\th{\theta}%
\global\long\def\l{\lambda}%

\global\long\def\Z{\mathcal{Z}}%
\global\long\def\z{\zeta}%
\global\long\def\ord#1{\mathcal{O}\left(#1\right)}%
\global\long\def\ua{\uparrow}%
\global\long\def\da{\downarrow}%
 
\global\long\def\co#1{\left[#1\right)}%
\global\long\def\oc#1{\left(#1\right]}%
\global\long\def\tr{\mbox{tr}}%
\global\long\def\o{\omega}%
\global\long\def\nab{\del}%
\global\long\def\p{\psi}%
\global\long\def\pro{\propto}%
\global\long\def\vf{\varphi}%
\global\long\def\f{\phi}%
\global\long\def\mark#1#2{\underset{#2}{\underbrace{#1}}}%
\global\long\def\markup#1#2{\overset{#2}{\overbrace{#1}}}%
\global\long\def\ra{\rightarrow}%
\global\long\def\cd{\cdot}%
\global\long\def\v#1{\vec{#1}}%
\global\long\def\fd#1#2{\frac{\d#1}{\d#2}}%
\global\long\def\P{\Psi}%
\global\long\def\dem{\overset{\mbox{!}}{=}}%
\global\long\def\Lam{\Lambda}%
 
\global\long\def\m{\mu}%
\global\long\def\n{\nu}%

\global\long\def\ul#1{\underline{#1}}%
\global\long\def\at#1#2{\biggl|_{#1}^{#2}}%
\global\long\def\lra{\leftrightarrow}%
\global\long\def\var{\mbox{var}}%
\global\long\def\E{\mathcal{E}}%
\global\long\def\Op#1#2#3#4#5{#1_{#4#5}^{#2#3}}%
\global\long\def\up#1#2{\overset{#2}{#1}}%
\global\long\def\down#1#2{\underset{#2}{#1}}%
\global\long\def\lb{\biggl[}%
\global\long\def\rb{\biggl]}%
\global\long\def\RG{\mathfrak{R}_{b}}%
\global\long\def\g{\gamma}%
\global\long\def\Ra{\Rightarrow}%
\global\long\def\x{\xi}%
\global\long\def\c{\chi}%
\global\long\def\res{\mbox{Res}}%
\global\long\def\dif{\mathbf{d}}%
\global\long\def\dd{\mathbf{d}}%
\global\long\def\grad{\vec{\del}}%

\global\long\def\mat#1#2#3#4{\left(\begin{array}{cc}
#1 & #2\\
#3 & #4
\end{array}\right)}%
\global\long\def\col#1#2{\left(\begin{array}{c}
#1\\
#2
\end{array}\right)}%
\global\long\def\sl#1{\cancel{#1}}%
\global\long\def\row#1#2{\left(\begin{array}{cc}
#1 & ,#2\end{array}\right)}%
\global\long\def\roww#1#2#3{\left(\begin{array}{ccc}
#1 & ,#2 & ,#3\end{array}\right)}%
\global\long\def\rowww#1#2#3#4{\left(\begin{array}{cccc}
#1 & ,#2 & ,#3 & ,#4\end{array}\right)}%
\global\long\def\matt#1#2#3#4#5#6#7#8#9{\left(\begin{array}{ccc}
#1 & #2 & #3\\
#4 & #5 & #6\\
#7 & #8 & #9
\end{array}\right)}%
\global\long\def\su{\uparrow}%
\global\long\def\sd{\downarrow}%
\global\long\def\coll#1#2#3{\left(\begin{array}{c}
#1\\
#2\\
#3
\end{array}\right)}%
\global\long\def\h#1{\hat{#1}}%
\global\long\def\colll#1#2#3#4{\left(\begin{array}{c}
#1\\
#2\\
#3\\
#4
\end{array}\right)}%
\global\long\def\check{\checked}%
\global\long\def\v#1{\vec{#1}}%
\global\long\def\S{\Sigma}%
\global\long\def\F{\Phi}%
\global\long\def\M{\mathcal{M}}%
\global\long\def\G{\Gamma}%
\global\long\def\im{\mbox{Im}}%
\global\long\def\til#1{\tilde{#1}}%
\global\long\def\kb{k_{B}}%
\global\long\def\k{\kappa}%
\global\long\def\ph{\phi}%
\global\long\def\el{\ell}%
\global\long\def\en{\mathcal{N}}%
\global\long\def\asy{\cong}%
\global\long\def\sbl{\biggl[}%
\global\long\def\sbr{\biggl]}%
\global\long\def\cbl{\biggl\{}%
\global\long\def\cbr{\biggl\}}%
\global\long\def\hg#1#2{\mbox{ }_{#1}F_{#2}}%
\global\long\def\J{\mathcal{J}}%
\global\long\def\diag#1{\mbox{diag}\left[#1\right]}%
\global\long\def\sign#1{\mbox{sgn}\left[#1\right]}%
\global\long\def\T{\th}%
\global\long\def\rp{\reals^{+}}%

\title{Attraction and condensation of driven tracers in a narrow channel}
\author{Asaf Miron$^{1}$, David Mukamel$^{1}$ and Harald A Posch$^{2}$}
\address{$\mbox{ }^{1}$Department of Physics of Complex Systems, Weizmann
Institute of Science, Rehovot 7610001, Israel}
\address{$\mbox{ }^{2}$Computational Physics Group, Faculty of Physics, University
of Vienna, Austria}

\begin{abstract}
Emergent bath-mediated attraction and condensation arise when multiple particles are simultaneously driven through an equilibrated bath under geometric constraints. While such scenarios are observed in a variety of non-equilibrium phenomena with an abundance of experimental and numerical evidence, little  \textit{quantitative} understanding of how these interactions arise is currently available. Here we approach the problem by studying the behavior of two driven "tracer" particles, propagating through a bath in a $1D$ lattice with excluded-volume interactions. We analytically explore the mechanism responsible for the tracers' emergent interactions and compute the resulting effective attractive potential. This mechanism is then numerically shown to extend to a realistic model of hard driven Brownian disks confined to a narrow $2D$ channel.
\end{abstract}
\maketitle

\section{Introduction \label{sec:Introduction}}

The physics governing the motion of a tracer particle, forcibly driven along a dense medium and confined to a narrow channel, has been extensively explored in recent decades. Besides its immediate relevance to various biological systems \citep{nestorovich2002designed,rout2003virtual,welte2004bidirectional,wente2010nuclear,kabachinski2015nuclear}, complex fluids, polymer solutions and glassy dynamics \citep{polin2008autocalibrated,gutsche2008colloids,kruger2009diffusion,Gazuz2009,candelier2010journey,dullens2011shear,Winter2012,Gazuz2013,Gruber2016}, it also lies at the core of two important technological applications: active microrheology, where microscopic probes are driven along a sample to investigate its spatio-temporal response properties \citep{Squires2005,Mizuno2008,Squires2010,wilson2011small}, and microfluidic devices, where colloids are manipulated through an intricate array of microscopic channels, carefully engineered to perform various tasks \citep{kirby2010micro,zhang2018particle}.

The ubiquity of driven tracer systems has stimulated vigorous theoretical activity aimed at providing understanding for the collective dynamics induced by a driven tracer in geometrically confined and strongly interacting many-body systems \citep{burlatsky1992directed,burlatsky1996motion,de1997dynamics,landim1998driven,benichou1999biased,illien2013active,Illien2014,cividini2016exact,cividini2016correlation,kundu2016exact,Leitmann2017,benichou2018unbinding,Miron_2020,miron2020driven}. While theoretical efforts have mostly focused on the case of a single driven tracer, the applications noted above typically involve \textit{multiple} simultaneously-driven tracers, substantially complicating the picture. Correspondingly, the cooperative behavior of multiple driven tracers has recently attracted considerable attention  \citep{mejia2011bias,vasilyev2017cooperative,Poncet2019,Lobaskin_2020, Kusters_2017}. Extensive numerical studies of multiple tracers in $2D$ and $3D$ systems have shown that the surrounding fluid mediates \textit{attractive} interactions between the tracers, leading them to cluster and form a condensate  \citep{mejia2011bias,vasilyev2017cooperative}. Studies of $1D$ lattice models whose single-file dynamics stems from a simple symmetric exclusion process (SSEP), whereby each lattice site may hold one particle at most, have also demonstrated cooperative aspects of the tracer dynamics \citep{Poncet2019,Lobaskin_2020}. In this setting, though, where particles cannot overtake each other, the tracers cannot get too close since the number of bath particles between each pair of tracers is conserved under the dynamics. In a recent numerical study involving a finite  density of driven tracers, it was shown that the tracers tend to aggregate and form clusters which, for a sufficiently narrow channel, could lead to the formation of a plug \citep{Kusters_2017}. Yet, in spite of the vast attention this problem has received, there is still no clear \textit{quantitative} understanding of the mechanism responsible for the bath-mediated attractive interactions between multiple tracers, nor their resulting condensation.
\begin{figure}
\begin{centering}
\includegraphics[scale=0.35]{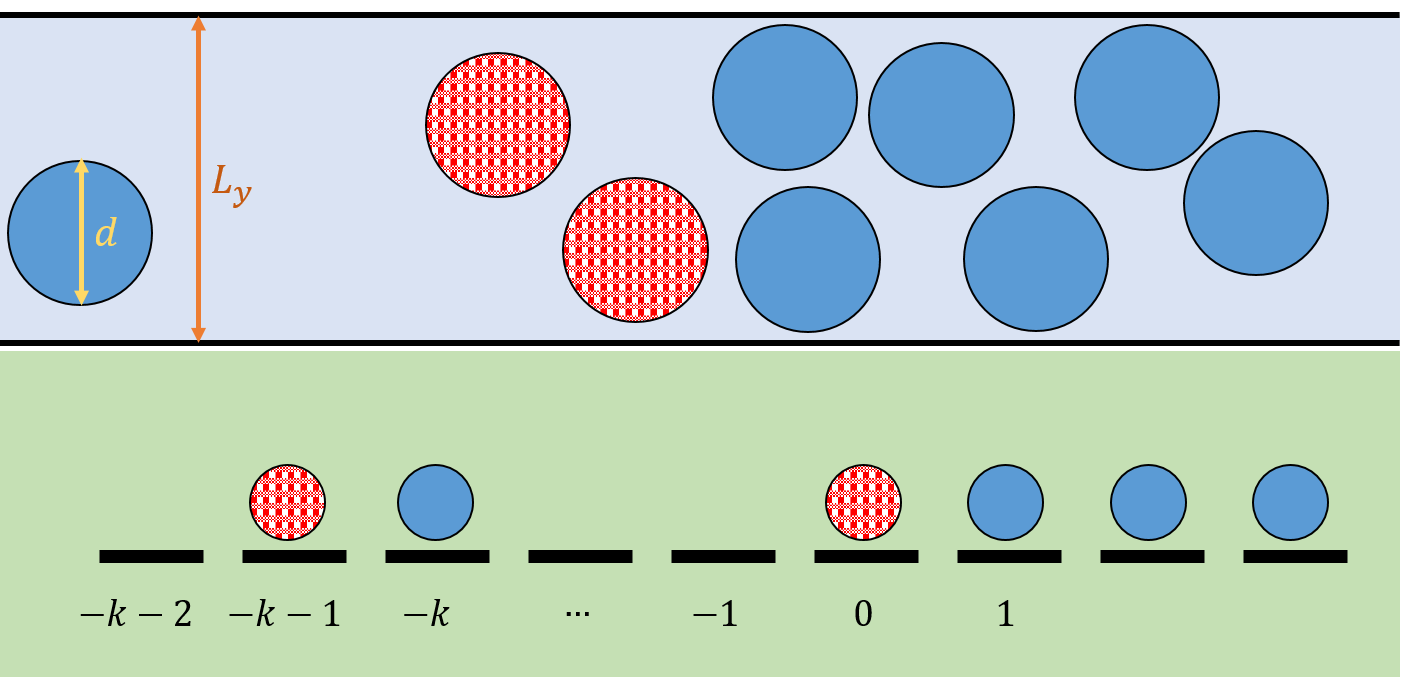}
\caption{Top: Schematic illustration of the narrow $2D$ channel system, which comprises of hard-core, disk-shaped Brownian particles of diameter $d$ inside a narrow channel of width $L_{y}>2d$. The driven tracers and bath particles are respectively depicted by checkered red disks and solid blue disks. 
Bottom: Schematic illustration of the corresponding $1D$ lattice model, explicitly demonstrating the site labeling appearing in Eqs. \eqref{eq:outrates}-\eqref{eq:inrates}.}
\label{fig:models}
\end{centering}
\end{figure}

\begin{figure}
\begin{centering}
\includegraphics[scale=0.615]{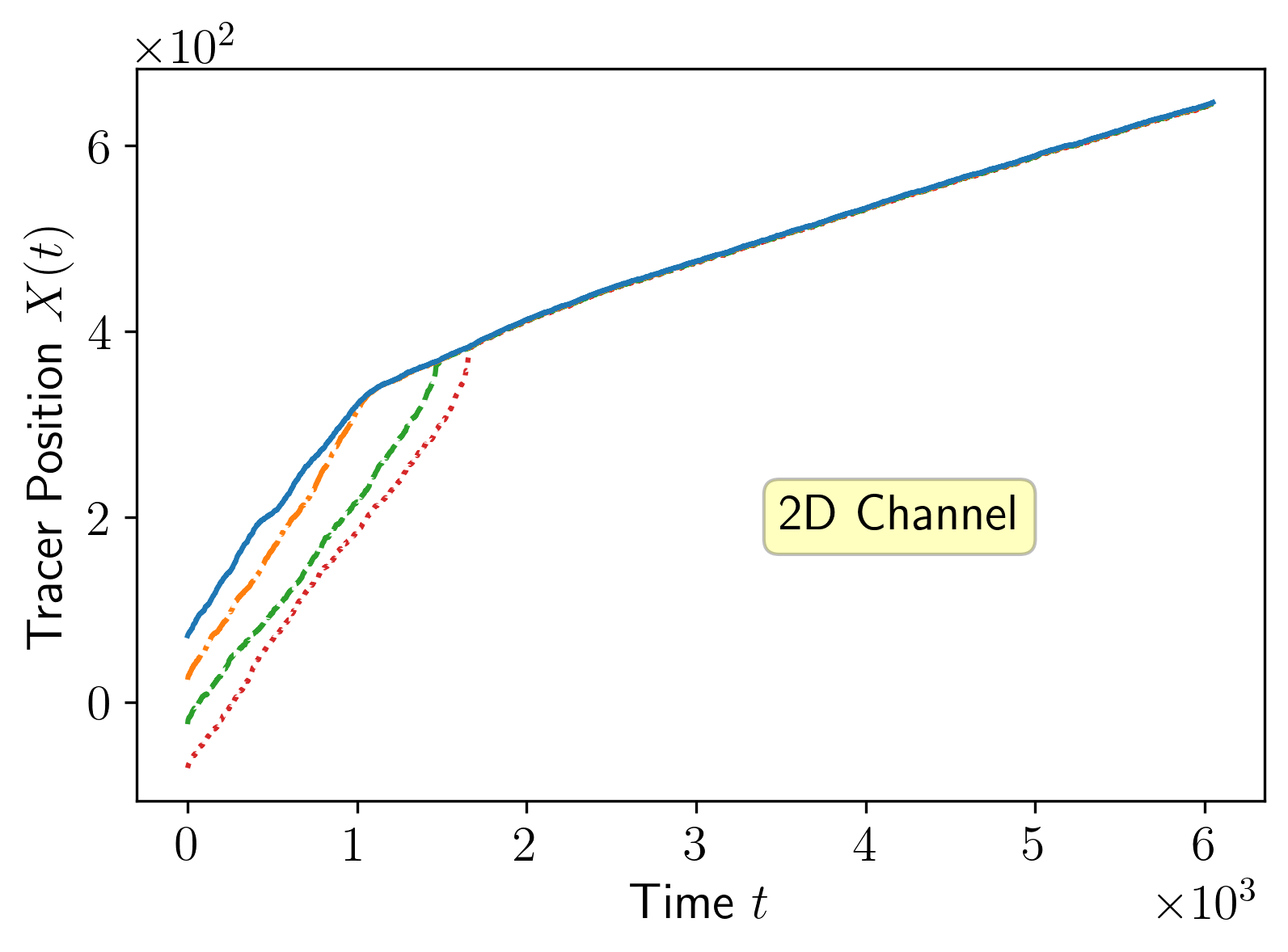}
\caption{Trajectories of four hard-core, Brownian disk tracers driven by a force $F=16$ through a bath of density $\bar{\r}=0.5$ along a $2D$ narrow channel of width $L_y=2.1$ and periodic length $L_x=190$. The model, and the reduced units in particular, are formally introduced in Sec.\ref{sec:2d}.}
\label{fig:channel_traj}
\end{centering}
\end{figure}

\begin{figure}
\begin{centering}
\includegraphics[scale=0.2]{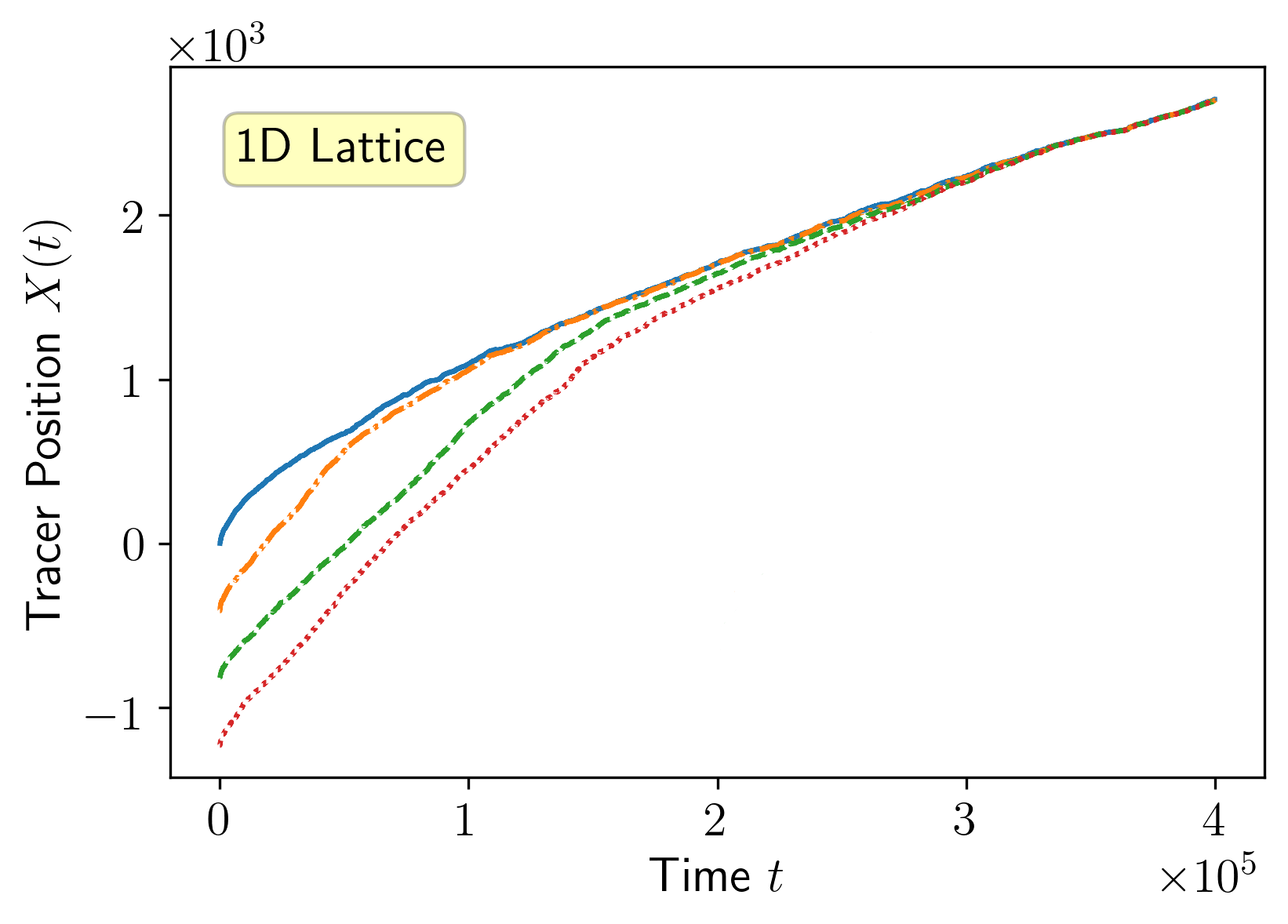}
\caption{Trajectories of four tracers in a ring of $L=2048$ sites with a bath density of $\bar{\r}=0.2$ and rates $p=1$ and $q=p'=q'=0.001$. }
\label{fig:lattice_traj} 
\end{centering}
\end{figure}

In this paper we study the mechanism responsible for the emergent attraction and condensation of multiple driven tracers in a narrow $2D$ channel, occupied by disk-shaped Brownian particles with hard-core interactions (See Fig. \ref{fig:models}). To this end we model the system as a $1D$ lattice SSEP, whose dynamics is extended to account for overtaking in the $2D$ channel. This approach is shown to capture the salient features of the attraction mechanism and has the advantage of being amenable to analytical treatment. Studying the $1D$ lattice model with two driven tracers, we uncover a phase in which the driven tracers strongly attract each other, forming a robust bound pair. This strong attraction can be attributed to an effective entropic potential between the tracers, shown to originate from the inhomogeneous bath density profile generated by the tracers, as seen in their respective co-moving reference frames. It acts to restrict their motion and mediate their emergent interactions. With this insight, we construct an effective framework in which the two driven tracers are modeled by independent biased random walkers, whose moving rates are determined by the local density of the surrounding bath. Within this framework, we compute the emergent attractive potential between the tracers and explicitly show how the bath's density forces the tracers towards one another, leading to the formation of a robust bound pair.
Extensive numerical simulations strongly suggest that the same underlying mechanism is at play in the narrow 2D channel (see Figs. \ref{fig:channel_traj} and \ref{fig:lattice_traj}). In particular, emergent tracer attraction and condensation are accompanied by qualitatively similar bath-density profiles and mean tracer velocities. Our results are, thus, argued to serve as a rather general mechanism for attraction and condensation in geometrically constrained systems involving multiple driven tracers.

The paper is structured as follows: Section \ref{sec:1d} provides a detailed account of the $1D$ lattice model and previous results that serve as the basis of the current study. It then presents our first main result, showing that the $1D$ lattice model features a phase in which the probability of finding two driven tracers at a distance $k$ decays exponentially with $k$. Section \ref{sec:2d} elaborates on the narrow $2D$ channel system and presents our second main result, showing that this setup exhibits a qualitatively similar mechanism for the attraction and condensation of multiple driven tracers. Section \ref{sec:conclusions} concludes the paper. Additional details and support of our results are provided in the Appendix.

\section{$1D$ Lattice Model \label{sec:1d}}

SSEP is a canonical lattice model for geometrically constrained and interacting many-body systems. It describes particles hopping along a $1D$ lattice with hard-core interactions that prevent particles from overtaking one another \citep{Mallick2015}. Yet we would like to consider settings where overtaking \textit{is} possible, albeit significantly suppressed by geometric constraints. We thus adopt an extension of the standard SSEP dynamics, first introduced in \citep{Miron_2020}, that includes driven "tracer" particles that can also overtake neighboring bath particles at a finite rate.

\begin{figure}
\begin{centering}
\includegraphics[scale=0.55]{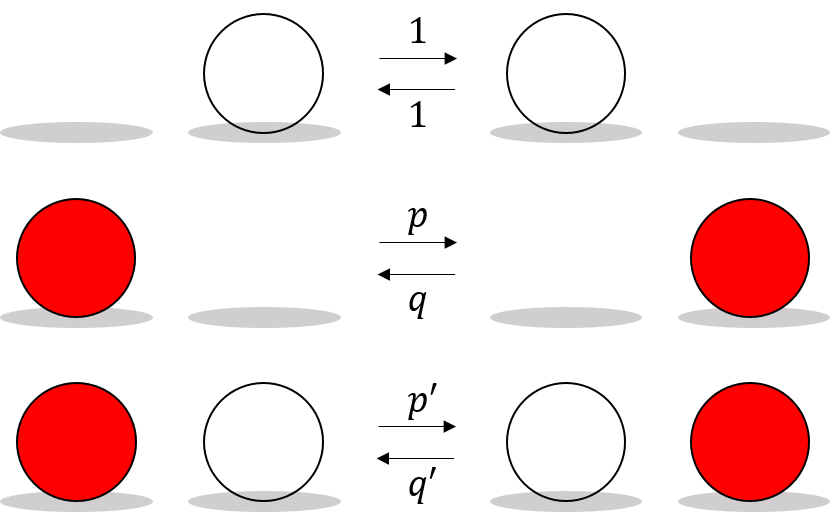} 
\par\end{centering}
\caption{Schematic illustration of the $1D$ lattice model dynamics. Bath particles are depicted by an empty circle and driven tracers are depicted by a red circle. Arrows represent the allowed moves with their respective attempt rates.}
\label{fig:illustration} 
\end{figure}

Consider a $1D$ ring of $L$ (even) sites, labeled $\el=-L/2+1,...,-1,0,1,...,L/2$, which are occupied by $M$ driven tracer particles and $N$ bath particles of mean density $\overline{\r}=N/(L-M)$. All particles interact via "simple exclusion" \citep{Mallick2015}, whereby each site may hold one particle at most. Bath particles symmetrically attempt hops to neighboring right and left sites  with rates $1$, whereas the tracers' hopping rates, $p$ to the right and $q$ to the left, are generally asymmetric with $p\ne q$. Particle overtaking is introduced into this dynamics by allowing a tracer to exchange sites with a bath particle occupying its neighboring right and left sites with respective rates $p'$ and $q'$ (see Fig. \ref{fig:illustration}).

\begin{figure}
\begin{centering}
\includegraphics[scale=0.6]{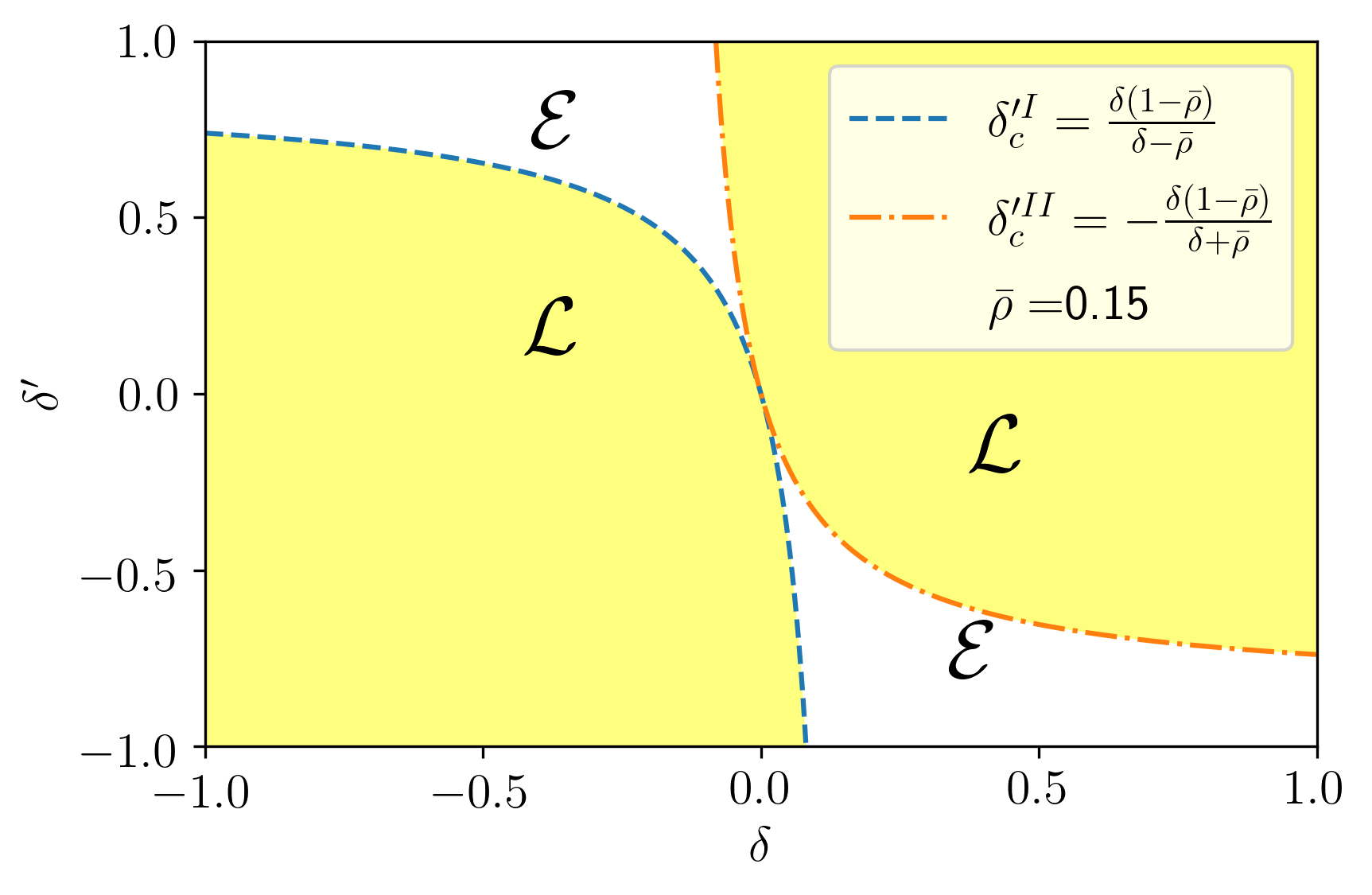}
\caption{The phase diagram for the $1D$ lattice model with a \textit{single} driven tracer and average bath density $\bar{\r}=0.15$. The localized phase appears in yellow while the extended phase is in white. The critical manifolds are given by $\d_{c}^{\prime I}=\frac{\d(1-\bar{\r})}{\d-\bar{\r}}$ and $\d_{c}^{\prime II}=-\frac{\d(1-\bar{\r})}{\d+\bar{\r}}$.}
\label{fig:phase_diagram_0}
\end{centering}
\end{figure}

This model's phase diagram -- presented for $\bar{\r}=0.15$ in Fig. \ref{fig:phase_diagram_0} after rewriting the dynamical rates according to  $p=r(1+\d)$, $q=r(1-\d)$, $p'=r'(1+\d')$, and $q'=r'(1-\d')$ -- was derived for a \textit{single} driven tracer in Ref. \citep{Miron_2020}.
With this notation, the single tracer phase diagram is expressed in terms of only three parameters: $\d=\frac{p-q}{p+q}$ - the hop bias, $\d'=\frac{p'-q'}{p'+q'}$ - the exchange bias and the average density $\bar{\r}$. The phase diagram is independent of the other two parameters: $r=\frac{p+q}{2}$, - the hop magnitude, and $r'=\frac{p'+q'}{2}$ - the exchange magnitude.

By varying $\d$, $\d'$ and $\bar{\r}$, two phases were demonstrated: a ``localized'' phase, where the driven tracer attains a finite velocity and generates a localized deviation of the bath density from $\bar{\r}$ (in its co-moving frame of reference), and an ``extended'' phase, where the tracer's velocity vanishes as the inverse system size and the bath density profile extends across the entire system.

Extensive numerical studies of the narrow $2D$ channel system with a single tracer, described in Sec. \ref{sec:2d}, reveal a clear correspondence between the characteristic features of the $1D$ lattice model's localized phase and those of the narrow $2D$ channel, when it is wide enough to allow particle overtaking.
In particular, it is found that the driven tracer velocity remains finite for large systems and that the bath density profile is mostly localized in front of the driven tracer. Here we are interested in the emergent attraction and condensation that arise when \textit{multiple} driven tracers are considered. The correspondence between the $1D$ model's localized phase and the narrow $2D$ channel for a \textit{single} driven tracer, suggest that the behavior of multiple tracers may also be probed within the $1D$ lattice model's localized phase. To this end, and as a first step towards probing the driven tracers' attraction mechanism, we consider the $1D$ lattice model with two driven tracers for parameters corresponding to the single-tracer localized phase. 

\begin{figure}
\begin{centering}
\includegraphics[scale=0.6]{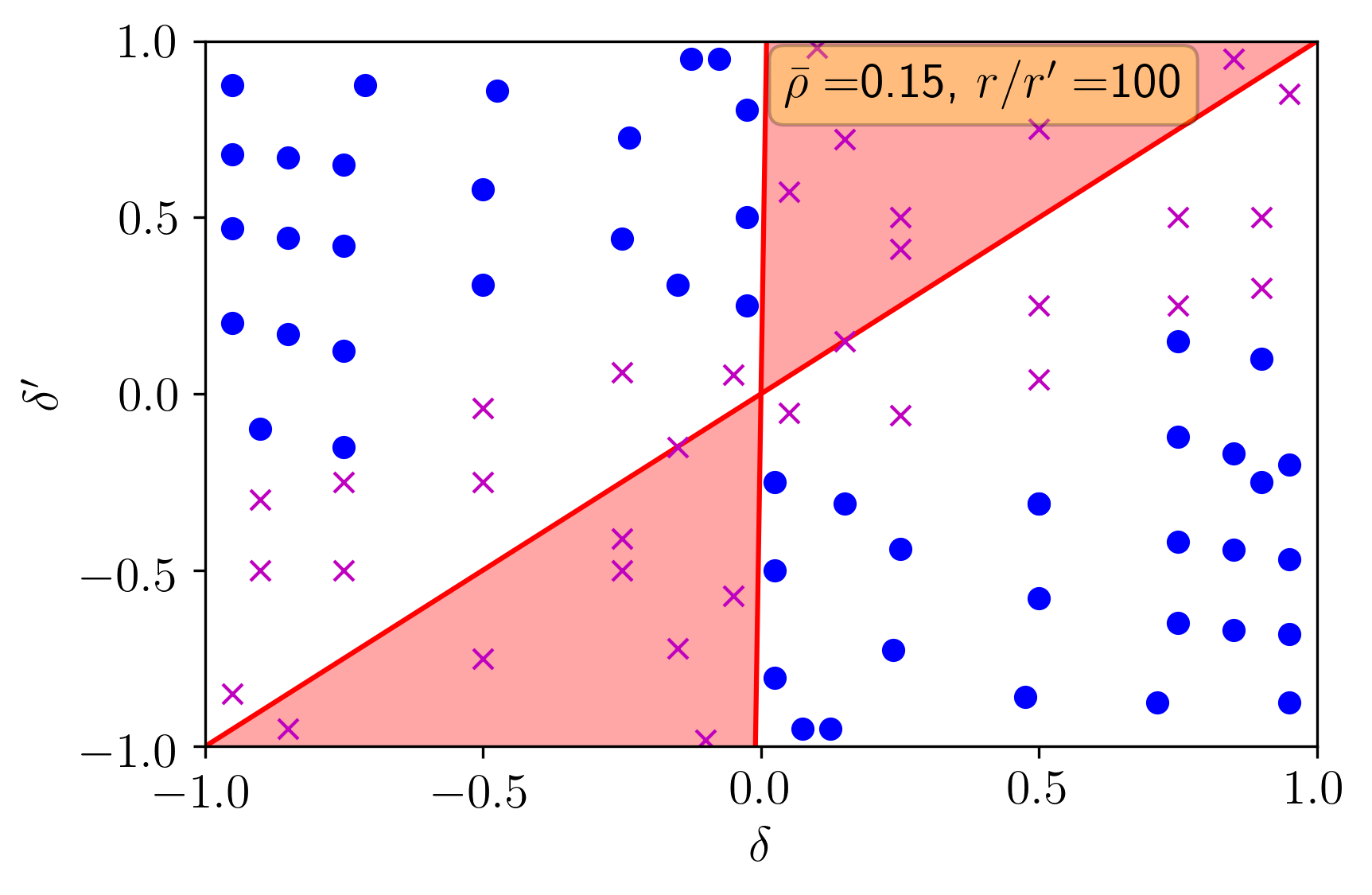}
\caption{The $1D$ lattice model's phase diagram for average bath density $\bar{\r}=0.15$ and $r/r'=100$. Repulsion is predicted in the red region while attraction is expected in the complementary domain. Data from numerical simulations of the model are illustrated by red crosses and blue dots, with the former depicting repulsion between the tracers and the latter depicting attraction.}
\label{fig:phase_diagram}
\end{centering}
\end{figure}

The resulting phase diagram in Figure \ref{fig:phase_diagram} shows that, in some domain of its parameter space, the $1D$ lattice model with two driven tracers features an "attractive" phase. In this phase the probability of finding the tracers separated by a distance $k$ is exponentially decaying. This is manifested in the tracers' tendency to strongly attract each-other and form a condensate.

\begin{figure}
\begin{centering}
\includegraphics[scale=0.58]{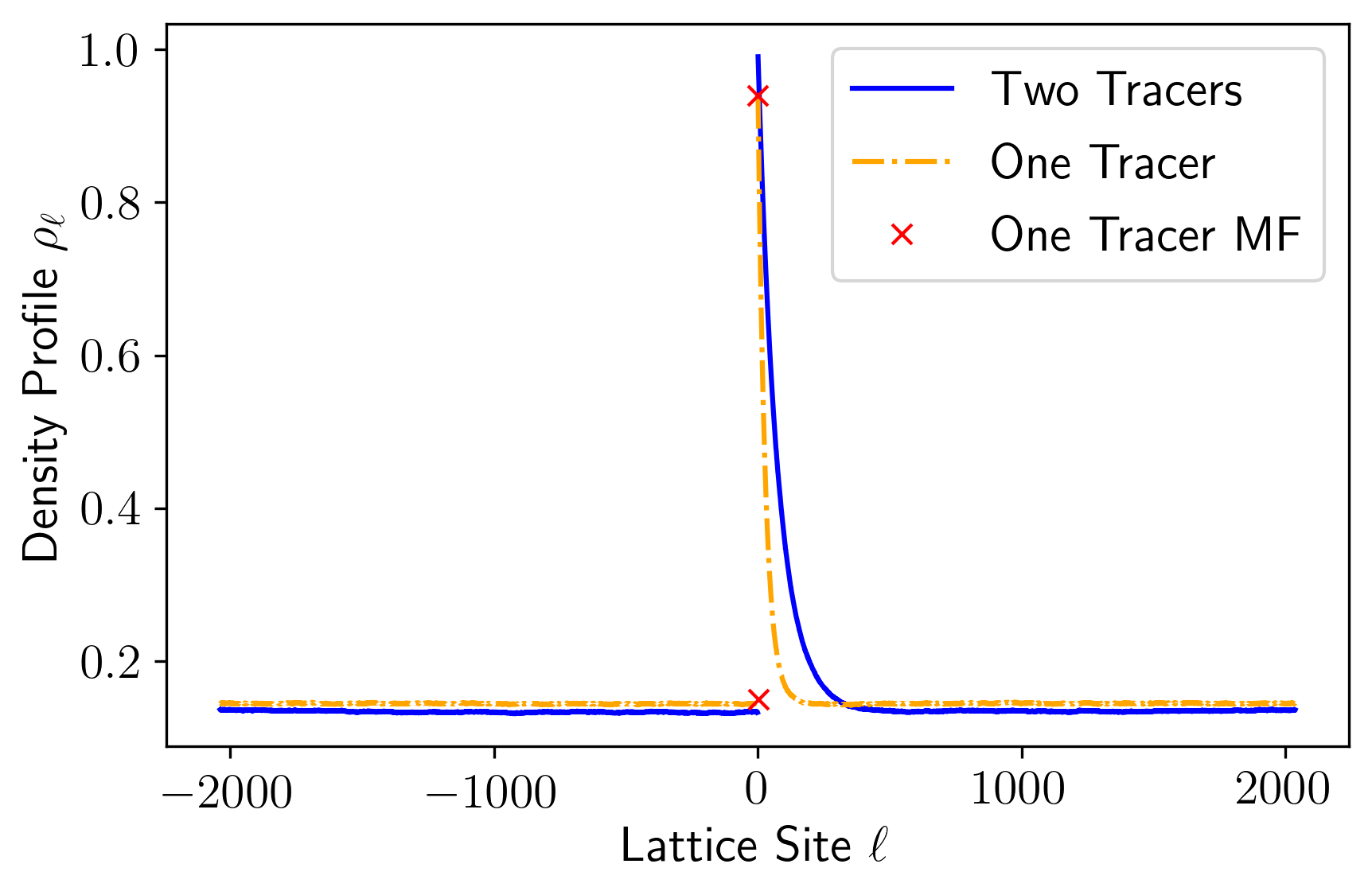}
\caption{The stationary bath density profile $\r_\el$ in a lattice of $L=4096$ sites with mean bath density $\bar{\r}=0.15$. The dynamical rates are $p=1.9$, $q=0.1$, $p'=0.0075$, and $q'=0.0125$, corresponding to $\d=0.9$, $\d'=-0.25$, $r=1$, and $r'=100$. The solid blue curve shows the bath density profile for two driven tracers, the dashed orange curve shows the profile for a \textit{single} driven tracer, and the red dots mark the single tracer mean-field prediction for the bath density adjacent to the tracers.
The deviation of the bath density near the bound tracer pair from the mean-field prediction for a \textit{single} driven tracer is small: at site $\el=1$ it is $\sim 0.0507$ and at site $\el=-1$ it is $\sim 0.0172$. The deviation of the simulated single tracer bath density profile from the mean field prediction is smaller still: at site $\el=1$ it is $\sim 0.0016$ and at site $\el=-1$ it is $\sim 0.0073$.}
\label{fig:lattice_den1}
\end{centering}
\end{figure}

This phase diagram is obtained within the mean-field approximation, where the two driven tracers are modeled as independent biased random walkers, whose moving rates depend on the stationary average bath density in their vicinity. To this end, we approximate the bath density profile with that derived in Ref. \citep{Miron_2020} for a \textit{single} driven tracer, which is shown in Fig. \ref{fig:lattice_den1} to be a fair approximation of the bath density profile generated by the bound tracer pair. We then demonstrate that, for strong drive magnitude $r$ and sufficiently small overtaking magnitude $r'$ (i.e. $r\gg r'$), the bath density acts to ''push'' the tracers towards one another. To quantify this effect, we take the thermodynamic limit $L\ra\infty$ and formulate an equation for the distribution $P_{n_{0},n_{1}}$ of the number of holes, $n_{0}$, and of bath particles, $n_{1}$, between the two tracers. We then solve $P_{n_{0},n_{1}}$ in the long-time limit and derive the stationary distribution $Q(k)$ of the distance $k=n_{0}+n_{1}$ between the tracers, which is given by $Q(k)=\sum_{n_{0}+n_{1}=k}P_{n_{0},n_{1}}$. Our analysis is concluded upon showing that $Q(k)$ decays exponentially with $k$, indicating that the tracers effectively interact via a linear attractive potential, which leads to the tracers' bound state.
Additional validation for the assumption that the bound tracer pair retains the features of the single-tracer localized phase is provided in Figs. \ref{fig:lattice_v} and \ref{fig:lattice_den} 
in the Appendix. 

As a first step in this analysis we consider a large ring, $L\gg1$, and take $n_0$ and $n_1$ whose sum corresponds to the short distance between the two tracers: 
the instantaneous position of the rightmost ``tracer 1'' \textit{defines} the location of site $\el=0$. The second tracer, called ``tracer 2'', is then located at site $\el=-(k+1)$ (see Fig. \ref{fig:models}). We henceforth refer to tracer moves that \textit{reduce} $k$ as "inward" moves, while moves that \textit{increase} $k$ are called "outward" moves.
The master equation for $P_{n_{0},n_{1}}$ is then
\[
\partial_{t}P_{n_{0},n_{1}}=\left(R_{+}^{h}\nabla^{n_{0}}_{-}+R_{-}^{h}\nabla^{n_{0}}_{+}\right)P_{n_{0},n_{1}}
\]
\begin{equation}
+\left(R_{+}^{x}\nabla^{n_{1}}_{-}+R_{-}^{x}\nabla^{n_{1}}_{+}\right)P_{n_{0},n_{1}},\label{eq:P_n0_n1}
\end{equation}
where $\nabla^{n_{0}}_{\pm}P_{n_{0},n_{1}}\equiv P_{n_{0}\pm1,n_{1}}-P_{n_{0},n_{1}}$ and $\nabla^{n_{1}}_{\pm}P_{n0,n1}\equiv P_{n_{0},n_{1}\pm1}-P_{n_{0},n_{1}}$ are discrete gradient operators. Equation \eqref{eq:P_n0_n1} is supplemented with the ''boundary'' condition $P_{n_{0},n_{1}}=0$ for $n_{0}<0$ or $n_{1}<0$.
The moving rates $R_{+}^{h},R_{-}^{h},R_{+}^{x},R_{-}^{x}$ respectively correspond to moves that introduce a hole between the tracers by hopping outwards, remove a hole by hopping inwards, introduce a bath particle by exchanging outwards, and remove a bath particle by exchanging inwards. 
Due to the model's exclusion interactions and exchange dynamics, the moving rates $R_{\pm}^{h,x}$ in Eq. \eqref{eq:P_n0_n1} naturally depend on the instantaneous bath particle occupation at the sites $\el=\mp1$ and $\el=-(k+1)\mp1$, which are adjacent to the locations of tracers 1 and 2, respectively.
Within the mean field approximation, we replace the instantaneous site occupations by their stationary average density. 
This completes the reduction of the model's dynamics to that of two independent biased random walkers, whose moving rates are determined by the stationary bath density around them. These rates are given by
\begin{align}
    R_{+}^{h} & =  p\left(1-\r_{1}\right)+q\left(1-\r_{-k-2}\right)~,~R_{+}^{x}=p'\r_{1}+q'\r_{-k-2}, 
    \label{eq:outrates}\\
    R_{-}^{h} & =  q\left(1-\r_{-1}\right)+p\left(1-\r_{-k}\right)~,~R_{-}^x=q'\r_{-1}+p'\r_{-k}, \label{eq:inrates}
\end{align}
with $\r_{\el}$ and $(1-\r_{\el})$ replacing the respective instantaneous occupation and vacancy of site $\el$. 

With the expressions provided by Eqs. \eqref{eq:P_n0_n1}-\eqref{eq:inrates}, we proceed to demonstrate the existence of an attractive phase in which the two tracers form a robust bound state. This is most easily understood after gaining some intuition by initially inspecting the most trivial setup:
Let us consider a forward-biased drive $p>q$ with $p\sim\ord{1}$ and note the following trivial observation: if the two tracers are initially placed at adjacent sites and no overtaking is allowed, i.e. $p'=q'=0$, the number of bath particles between the tracers remains zero at all future times. The two tracers cannot get too far from each other, as this would require an unreasonably large fluctuation of the density of bath particles. Thus, the two tracers form a bound state. 
While the existence of a two-tracer bound state is not very surprising under such geometrically restrictive settings, its persistence in the presence of \textit{finite} overtaking rates is far from obvious.

Let us now introduce small overtaking rates $p', q' \ll 1$ (equivalently $r/r'\gg1$), and analyze the distance distribution between the two tracers. Determining the tracers' moving rates, $R_{\pm}^{h},R_{\pm}^{x}$ in Eqs. \eqref{eq:outrates} and \eqref{eq:inrates} requires knowledge of the stationary bath densities near the tracers. Motivated by the equivalence between the $1D$ lattice model's localized phase and the narrow $2D$ channel model for a single driven tracer, we proceed to carry out the analysis in the localized phase of the $1D$ lattice model. If the two tracers do form a bound state, where the typical distance between them is small (of $\sim\ord{1}$ with respect to $L$), we may think of this bound pair as a single ``effective'' driven tracer. The bath density profile $\r_\el$ obtained in Ref. \citep{Miron_2020} for a \textit{single} driven tracer can then be used to approximate that generated by the bound pair (see Fig. \ref{fig:lattice_den1}).
We thus use the results 
\begin{equation}
    \r_{1}=\overline{\r}\frac{p-\left(1-\overline{\r}\right)\left(q-q'\right)}{p\overline{\r}+p'\left(1-\overline{\r}\right)} ~,\label{eq:rho_1 localized}
\end{equation}
and
\begin{equation}
    \r_{-(k+2)}=\overline{\r} ~, \label{eq:rho_-1 localized}
\end{equation}
obtained in Ref. \citep{Miron_2020} for the localized phase to determine the outward moving rates $R_+^{h}$ and $R_+^{x}$ given in Eq. (\ref{eq:outrates}). We are then left to determine the inwards moving rates $R_-^h,R_-^x$. 


Let us now consider the limit of a large forward drive, $p\sim O(1)$ and $q\ll 1$, which together with small overtaking rates $p', q' \ll 1$ result in small \textit{outward} rates $R_{+}^{h},R_{+}^{x}$.
On time scales smaller than the inverse of $R_{+}^{h},R_{+}^{x}$, the distance between the tracers remains mostly unchanged. The bath particles located between them, which symmetrically attempt hops to both sides with rate $1$, quickly relax to a stationary state, which in the case of SSEP in a closed interval is a uniform distribution. We thus approximate the distribution of the particles between the two tracers by a uniform distribution, implying $\r_{-1}=\r_{-k}=\frac{n_{1}}{n_{0}+n_{1}}$ and $1-\r_{-1}=1-\r_{-k}=\frac{n_{0}}{n_{0}+n_{1}}$. 
For an $(n_0,n_1)$ configuration, the inward rates in Eq. \eqref{eq:inrates} are 

\[
R^{h}_{-}(n_0,n_1)=(p+q)\frac{n_0}{n_0+n_1},
\]
\begin{equation}
R^{x}_{-}(n_0,n_1)=(p'+q')\frac{n_1}{n_0+n_1}. \label{eq:inrates hop+ex}
\end{equation}

Equations \eqref{eq:P_n0_n1}-\eqref{eq:inrates hop+ex}
can now be used to determine the stationary distribution $P_{n_0,n_1}$.
Verification of the two-tracer bound state in the \textit{absence} of overtaking provides a welcome consistency check. Substituting $n_{1}=p'=q'=0$ yields $P_{n_{0},0}\pro\left(R_{+}^{h}/(p+q)\right)^{n_{0}}$. Since $R_{+}^h/(p+q)\ll1$, the stationary probability indeed decays exponentially with $n_0$. 

We are finally ready to tackle the far more interesting case of \textit{finite} exchange rates $p',q'\ll1$. We seek an exponentially decaying solution which generalizes the above solution in the absence of overtaking. This motivates the ansatz  
\begin{equation}
    P_{n_0,n_1}=A_{n_0,n_1} a^{n_0}b^{n_1},
    \label{eq:P(a,b)}
\end{equation}
where
\begin{equation}
    a\equiv R_+^h/(p+q), b\equiv R_+^x/(p'+q'),
    \label{eq:a,b}
\end{equation}
and $A_{n_0,n_1}$ remain to be determined. 
Inserting the ansatz in Eq. \eqref{eq:P(a,b)} into Eq. \eqref{eq:P_n0_n1} yields 


\[
(p+q)\left(A_{n_0-1,n_1}-\frac{n_0}{n_0+n_1}A_{n_0,n_1}\right)\]
\[
+(p'+q')\left(A_{n_0,n_1-1}-\frac{n_1}{n_0+n_1}A_{n_0,n_1}\right)
\]
\[
+R^{h}_{+}\left(\frac{n_1+1}{n_0+n_1+1}A_{n_0,n_1+1}-A_{n_0,n_1}\right)
\]
\begin{equation}
+R^{x}_{+}\left(\frac{n_0+1}{n_0+n_1+1}A_{n_0+1,n_1}-A_{n_0,n_1}\right)=0,
    \label{eq:A(n_0,n_1)}
\end{equation}
for $n_0,n_1>0$. Each pair of brackets in Eq. \eqref{eq:A(n_0,n_1)} individually vanishes for $A_{n_0,n_1} = A{{n_0+n_1}\choose{n_0}}$, with $A\equiv A(n_0,0)=A(0,n_1)$ set by the normalization of $P_{n_{0},n_{1}}$, giving
\begin{equation}
P_{n_{0},n_{1}}=\left(1-a-b\right){n_{0}+n_{1} \choose n_{0}}a^{n_{0}}b^{n_{1}}. \label{eq:P(a,b)final}
\end{equation}
Having found the distribution of holes, $n_{0}$, and of bath particles, $n_{1}$, between the tracers, we finally carry out the last step of our analysis. Taking a partial sum of $P_{n_{0},n_{1}}$ while keeping the distance between the two tracers $k\equiv n_0+n_1$ fixed yields 
\begin{equation}
Q(k)\equiv\sum_{n_{0}+n_{1}=k}P_{n_{0},n_{1}}=(1-a-b)(a+b)^k.\label{eq:Q_k}
\end{equation}

In analogy with equilibrium statistical mechanics, where the Boltzmann distribution for a configuration of energy $\E$ is $\sim e^{-\beta \E}$ at inverse temperature $\beta$, the exponential decay of the probability $Q(k)$ with $k$ can be attributed to a strongly attractive emergent potential 
\begin{equation}
\b \E=\b V(k)=-\ln({a+b)}k. 
\label{eq:potential}
\end{equation}

\begin{figure}
\begin{centering}
\includegraphics[scale=0.58]{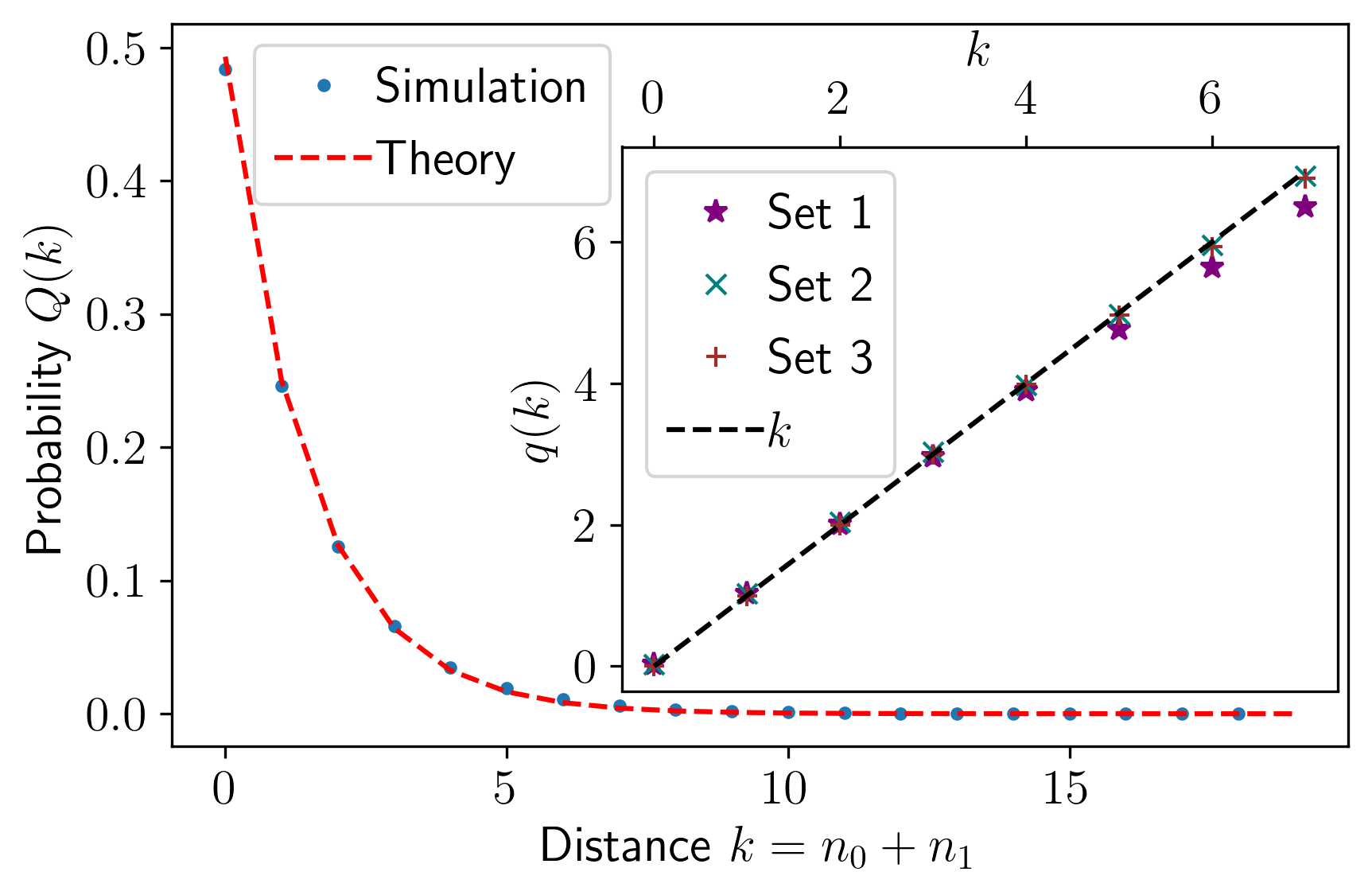}
\caption{Histogram $Q(k)$ of the distance $k=n_0+n_1$ between the two tracers in the $1D$ lattice ring of $L=4096$ sites with mean bath density $\bar{\r}=0.15$. Blue dots denote simulation data for $p=1.9$, $q=0.1$, $p'=0.0075$ and $q'=0.0125$, corresponding to $\delta=0.9$, $\delta'=-0.25$, $r=1$ and $r'=0.01$. The dashed red curve is the theoretical prediction for $Q(k)$ in Eq.\eqref{eq:Q_k}, with $R^{h}_{+}$ in $a$ and $R^{x}_{+}$ in $b$ have been determined by using the empirical bath density near the bound tracer pair. The inset shows a plot of $q(k)=\log{[\frac{Q(k)}{1-a-b}]}/\log{[a+b]}$ versus $k$ for three different sets of parameters $\delta$ and $\delta'$, for fixed $r=1$ and $r'=0.01$. Set 1: $\delta=0.9$, $\delta'=-0.25$. Set 2: $\delta=0.5$, $\delta'=-0.31$. Set 3: $\delta=0.75$, $\delta'=-0.65$.}
\label{fig:Q_k_fig}
\end{centering}
\end{figure}

The origin of this confining potential, which leads to a two-tracer bound state whose average distance is $\left <k\right>=(a+b)/(1-a-b)\ll1$, lies in the non-homogeneous bath density profile generated by the driven tracers themselves.
Equation \eqref{eq:Q_k} clearly relates the bath density profile near the tracers, which enters via $a$ and $b$ of Eqs. \eqref{eq:a,b}, to their emergent attractive interactions and resulting condensation. In Fig. \ref{fig:Q_k_fig} we compare the theoretical expression for $Q(k)$ in Eq. \eqref{eq:Q_k} to numerical simulation results, showing a very good agreement for several choices of the dynamical rates.

As is evident from the numerical two-tracer phase diagram presented in Fig. \ref{fig:phase_diagram}, attraction and condensation reach far beyond the parameter region corresponding to the single-tracer localized phase and appears to cover the entire extended phase as well (see Fig. \ref{fig:phase_diagram_0}). While our analysis was focused on the $1D$ lattice model's localized phase, in hopes of capturing the corresponding physics of a narrow $2D$ channel, this analysis is straightforward to apply to the extended phase too. Carrying out the analysis shows that the empirically observed attraction in the extended phase agrees with the results of our theoretical analysis. As in the localized phase, the analysis holds under the assumption that the bath density profile generated by the bound two-tracer pair is well approximated by the profile generated by a \textit{single} driven tracer in the extended phase. Substituting $\r_{1}=\frac{q'(p-q)}{pq'-qp'}$ and $\r_{-(k+2)}=\frac{q'(p-q)}{pq'-qp'}$, which were computed for the extended phase in Ref. \citep{Miron_2020}, inside $(a+b)^{k}$ of Eq. \eqref{eq:Q_k} for $Q(k)$ yields
\begin{equation}
Q(k)\pro\left(\frac{R^{h}_{+}}{p+q} + \frac{R^{x}_{+}}{p'+q'}\right)^{k}=(1+\d\d')^k.\label{eq:a+b_extended}
\end{equation}
Noting that $\d\d'<1$ is satisfied throughout the \textit{entire} extended phase \citep{Miron_2020}, we confirm that the distance between the two tracers $Q(k)\pro (a+b)^k$ indeed decays exponentially throughout the extended phase.

The phase diagram resulting from our analysis is given in Fig. \ref{fig:phase_diagram}. It displays a phase, in which the two tracers attract each other for $a+b<1$, and a repulsive phase in the complementary region. The phase diagram is compared with results obtained by numerical simulations of the model where the attractive and repulsive regions are indicated by blue dots and red x's, respectively. It is clearly seen that our analysis captures well the bulk of the two phases. Finding the transition line requires a more detailed analysis. This is expected since the assumption of a homogeneous distribution of the bath particles between the two tracers  is expected to work well for $a,b\ll 1$ but should break down at larger values of these parameters, where the transition is expected.


\section{2D Narrow Channel \label{sec:2d}}

In this section we provide details regarding the $2D$ narrow channel system and discuss its numerical simulation scheme. We then present numerical evidence showing that this system exhibits the same qualitative features found in the $1D$ lattice model's localized phase for a \textit{single} driven tracer. The bath density profile is \textit{localized} around the tracer, whose velocity approaches a \textit{finite} value at large system sizes. Finally, we show that this localized phase persists in the case of two driven tracers, and verify that the distribution of the distance between the tracers decays exponentially, just as it does in some regions of the $1D$ lattice model's localized phase. 
This evidence strongly suggests that the mechanism responsible for the attraction and condensation of multiple driven tracers in the narrow $2D$ channel is the same as that shown analytically in the $1D$ lattice model's localized phase. Namely, that the inhomogeneous bath density profile generated by the driven tracers conspires to induce a strong effective attraction between them. 

The narrow $2D$ channel system consists of $N+M$ disk-shaped particles of diameter $d$ and unit mass, including $N$ ''bath'' disks and $M$ ''tracer'' disks. The mean bath density is defined by $\bar{\r}=\frac{N}{L_{x}L_{y}}$, where $L_{x}$ is the channel's length and $2d<L_{y}\ll L_{x}$ its width. The channel is periodic along the $x$ direction.

The tracers are externally driven by a constant force of magnitude $F$ along the positive $x$ direction. The disks' positions and velocities, respectively denoted by $\boldsymbol{r}_{i}=\left(x_{i},y_{i}\right)$ and $\boldsymbol{v}_{i}=\left(v_{x,i},v_{y,i}\right)$ with $i=1,...,N+M$, evolve according to the following Langevin dynamics

\begin{equation}
\dot{\boldsymbol{r}}_{i}=\boldsymbol{v}_{i}\text{ ; }\dot{\boldsymbol{v}}_{i}=-\g\boldsymbol{v}_{i}+\sqrt{2\g\kb T}\boldsymbol{\xi}_{i}+ \left(\boldsymbol{F}\right)+\left\{ coll\right\}, \label{eq:channel_EOM}
\end{equation}
where $\g$ denotes the friction coefficient, $\kb$ the Boltzmann constant, $T$ is the temperature, $\boldsymbol{\xi}_{i}\left(t\right)$ is a $2D$ vector whose components are Gaussian white noises satisfying $\left\langle\xi_{i}^{\m}\left(t\right)\xi_{j}^{\n}\left(t'\right)\right\rangle =\d_{ij}\d^{\m\n}\d\left(t-t'\right)$, $\d_{ij}$ and $\d^{\m\n}$ are Kronecker delta functions, $i,j={1,2,...,N+M}$ denote particle indices, $\m,\n=1,2$ denote the two spatial directions, $\d\left(t-t'\right)$ is the Dirac delta function, and the notation $\left(\boldsymbol{F}\right)$ is used to indicate that the constant force $\boldsymbol{F}=F\hat{x}$ acts \textit{only} on the $M$ tracers and does not affect the $N$ bath disks. The last term in Eq. \eqref{eq:channel_EOM} encapsulates two kinds of collisions: elastic particle-particle collisions and ''thermal'' particle-wall collisions. In the latter case, the re-emitted disk's velocity components parallel and perpendicular to the wall are drawn from two distinct distributions: the Boltzmann distribution for the parallel component and a different distribution for the perpendicular one (See \citep{Tehver1998} for details). Dimensionless units are used, for which the disk diameter $d$, mass $m$ as well as $k_B$ and $T$, are all unity. In addition, $\g = 2$ is used throughout. 
The stochastic equations of motion are solved to first order in the time step $\Delta t = 0.001$ 
according to the Gillespie updating scheme 
\citep{Gillespie1,Gillespie2}.

In the following, the linear density profile $n(r)$ will be required. It may be approximated by covering the channel with a linear grid of boxes co-moving with the frame of one of the tracers which we refer to as ''tracer 1''. The width of the boxes is $L_{y}$, their length $\Delta x$. The latter may be chosen as required, $\Delta x \approx (N+M-1)/(4 L_x) \approx 0.25$ is a useful choice. The $\approx$ sign is required here to insure that $L_x/\Delta x$, the number of boxes,  is an integer. Let $N_{i}$ denote the \textit{total} number of particles in box $i$, including the second tracer if present (not to be confused with the number of bath particles $N$). This box is located at position $r_{i} = x_i - x_{t,1}$ (i.e. in tracer 1's frame of reference) where $x_{i}= i\Delta x$ and $i=0,...,L_{x}/\Delta x-1$. We then define the linear density profile $ n(r_i) = \langle (N+M-1) N_{i} / \sum_j N_j \Delta x \rangle$, where 
$\langle \dots \rangle $ denotes a time average. 
Unless noted otherwise, all figures in this section are obtained for parameters $F = 16$, $\bar{\r} = 0.4$ and $L_y = 2.6$, to which we hereafter refer to as the ''canonical'' set of parameters.

\subsection{Single driven tracer}
\begin{figure}
\begin{center}
\includegraphics[width=0.48\textwidth]{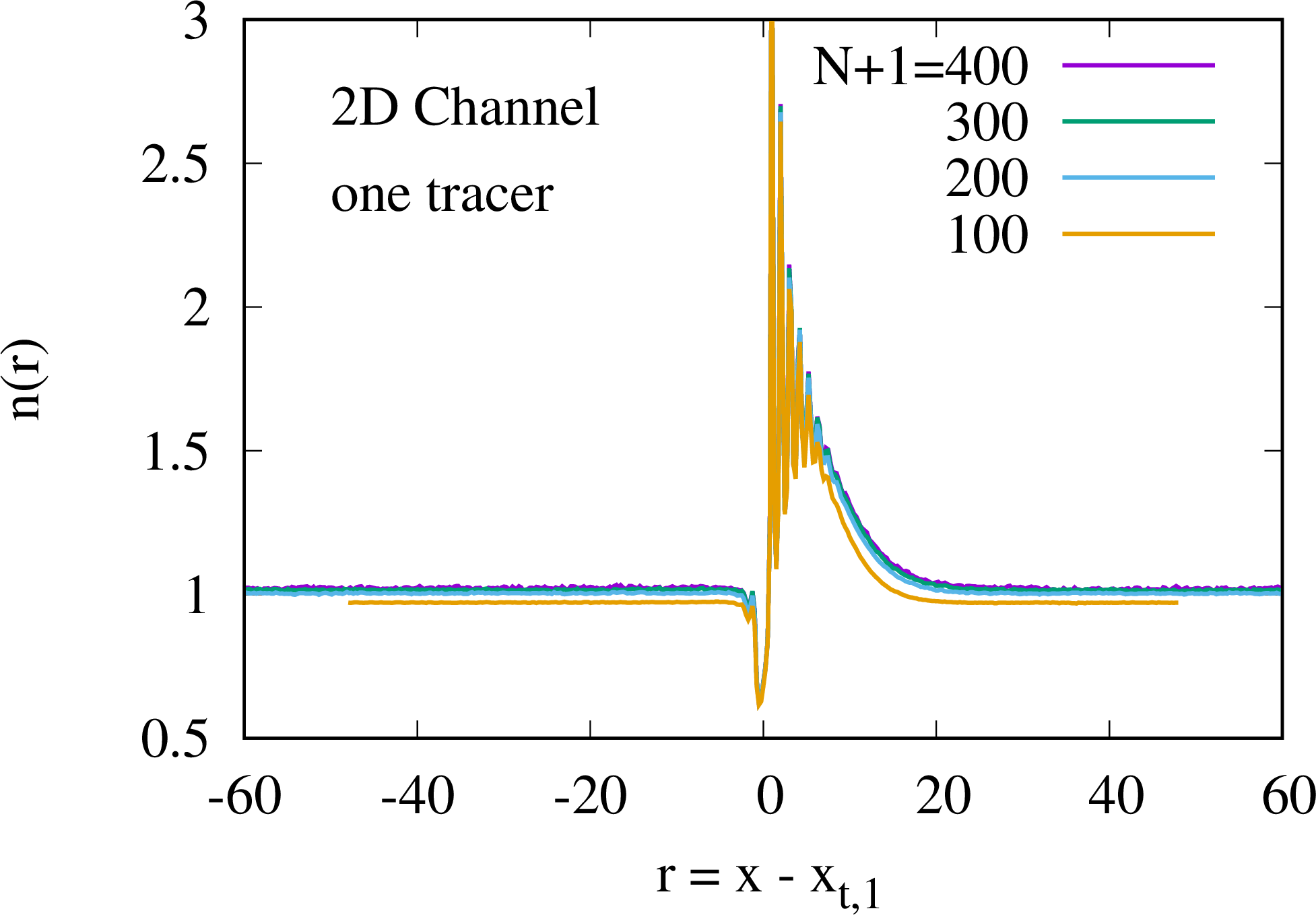}
\caption{System size dependence of the density profile for the $2D$ narrow channel system for the canonical parameters and a single driven tracer. Four profiles for various $L_x$ (or, equivalently, $N$) are shown, demonstrating data collapse for large $L_x$.}
\label{n_one_tracer_channel}
\end{center}
\end{figure}

We next demonstrate the qualitative similarities between the $1D$ lattice model's localized phase and the narrow $2D$ channel system for the case of a \textit{single} driven tracer. Extensive numerical studies  for a broad range of the parameters $F, \bar{\r}$ and $L_y$ show that, in the long time limit, the system reaches a stationary state with the characteristic features of the localized state: it exhibits a localized density profile, as seen in the tracer's reference frame, and the tracer's velocity reaches a finite value in the limit of large $L_x$. This is demonstrated in Fig. \ref{v_L_channel}, which displays the tracer's velocity as a function of $L_x$, and in Fig. \ref{n_one_tracer_channel}, where a data collapse of the bath density profile, $n(r)$, is shown for several values of $L_x$. 
The oscillations near the origin are a consequence of the crystal-like spatial arrangement of the dense bath particles that accumulate in front of the tracer. 

\subsection{Multiple driven tracers}
\begin{figure}
\begin{center}
\includegraphics[width=0.5\textwidth]{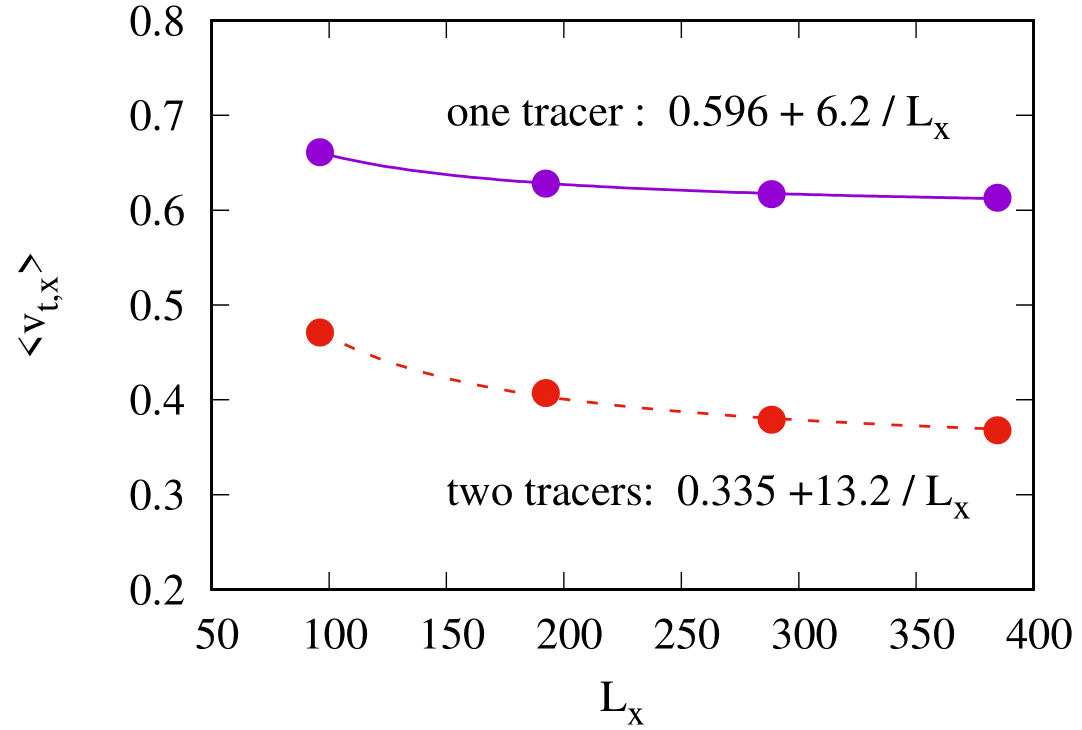}
\caption{Stationary mean tracer velocity $\left< v_{t,x}\right >$ along the channel axis as a function of the channel length $L_x$ in the $2D$ channel. The violet and red solid points respectively show numerical simulation data for a single tracer and a bound pair. The corresponding curves are a fit to the $a+b/L_x$ behavior that is expected in the localized phase. Canonical parameters, $F = 16$, $\bar{\r} = 0.4$ and $L_y = 2.6$, apply.}
\label{v_L_channel}
\end{center}
\end{figure}

\begin{figure}
\begin{center}
\includegraphics[width=0.48\textwidth]{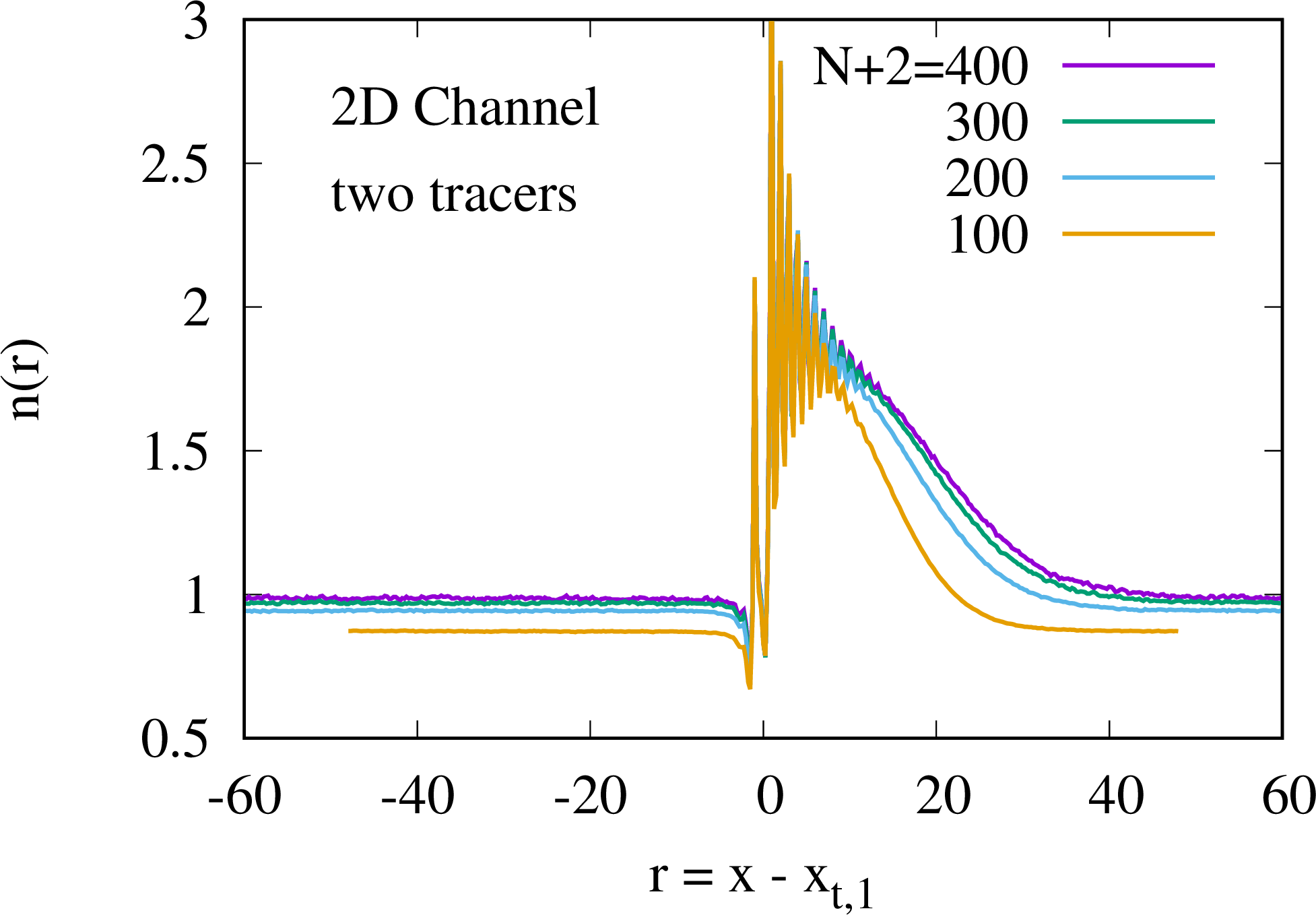}
\caption{System size dependence of the density profile for the narrow $2D$ channel system for the canonical parameters and two driven tracers. Four curves are plotted for different values of $L_x$ (or, equivalently, $N$). For the longest channels a data collapse takes place.}
\label{n_two_tracers_channel}
\end{center}
\end{figure}

We now turn to the case of multiple tracers. 
Figure \ref{fig:channel_traj} shows typical trajectories of four tracers, clearly demonstrating the formation of a condensate in the long-time limit.  
Focusing on the case of two tracers, we demonstrate in Figs. \ref{v_L_channel} and  \ref{n_two_tracers_channel} that in analogy with the $1D$ lattice model, the channel system retains the localized phase's characteristic features: i) a density profile $n(r)$ that is concentrated near the bound tracer pair and ii) a non-vanishing velocity of the bound tracer pair in the large $L_x$ limit. Analogous plots for the $1D$ lattice model are shown in the Appendix Figs. \ref{fig:lattice_den} and \ref{fig:lattice_v}.

Figure \ref{n_two_tracers_channel} shows a
data collapse, for various $L_x$, of the stationary density profiles $n(r)$ for two driven tracers. As commented above for the case of single tracer, the oscillations near the origin are a consequence of the crystal-like spatial arrangement of the dense bath particles that accumulate in front of the tracers.
The localized two tracer density profile is analogous to the one appearing in Fig. \ref{fig:lattice_den} for the $1D$ lattice model. 

\begin{figure}
\begin{center}
\includegraphics[width=0.475\textwidth]{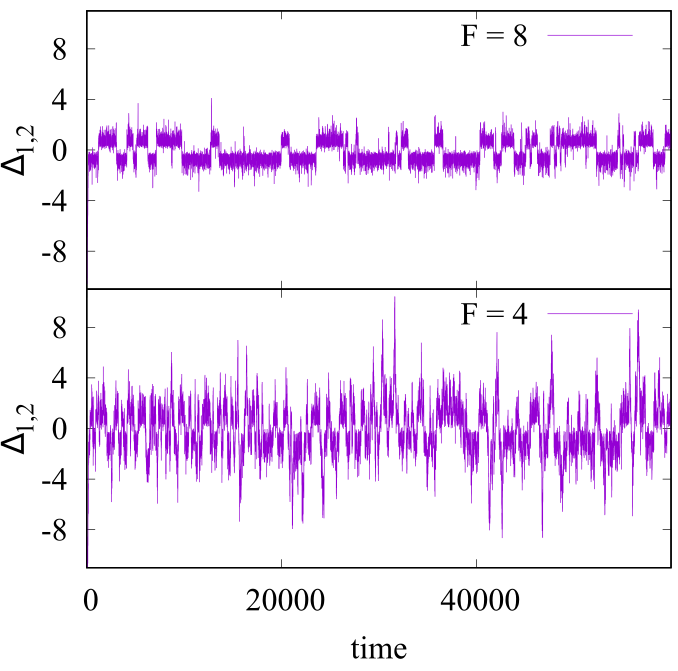}
\caption{Separation of the two tracers in a narrow channel of width $L_y = 2.1$, average density $\bar{\r}=0.5$ and 
$L_x = 190.5$ as a function of time.
Top: $F = 8$. This field is strong enough to
prevent the particle pair from separating.
Bottom: $F = 4$. This value is close to the
separation threshold of the two tracers. 
}
\label{delta_channel}
\end{center}
\end{figure}

The strong attraction between the tracers is demonstrated in Fig. \ref{delta_channel}, which displays the $x$-separation $\D_{1,2}$ between the tracers as a function of time. Two different values of the driving force are shown for channel width $L_y=2.1$ and mean bath density $\bar{\r} = 0.5$. 
In the top panel, obtained for $F=8$, the distance is shown to remain very small at all times, and appears to be roughly of $\sim\ord{1}$. The bottom panel, obtained for $F=4$, demonstrates that this typical distance grows as the driving force is weakened. Further reducing $F$ beyond some critical value, or equivalently, increasing the channel width $L_{y}$ at fixed $F$, breaks the tracers' bound state. Nevertheless, this bound state is observed to persist for a range of $F$ and $L_{y}$.

\begin{figure}
\begin{center}
\includegraphics[width=
0.48\textwidth]{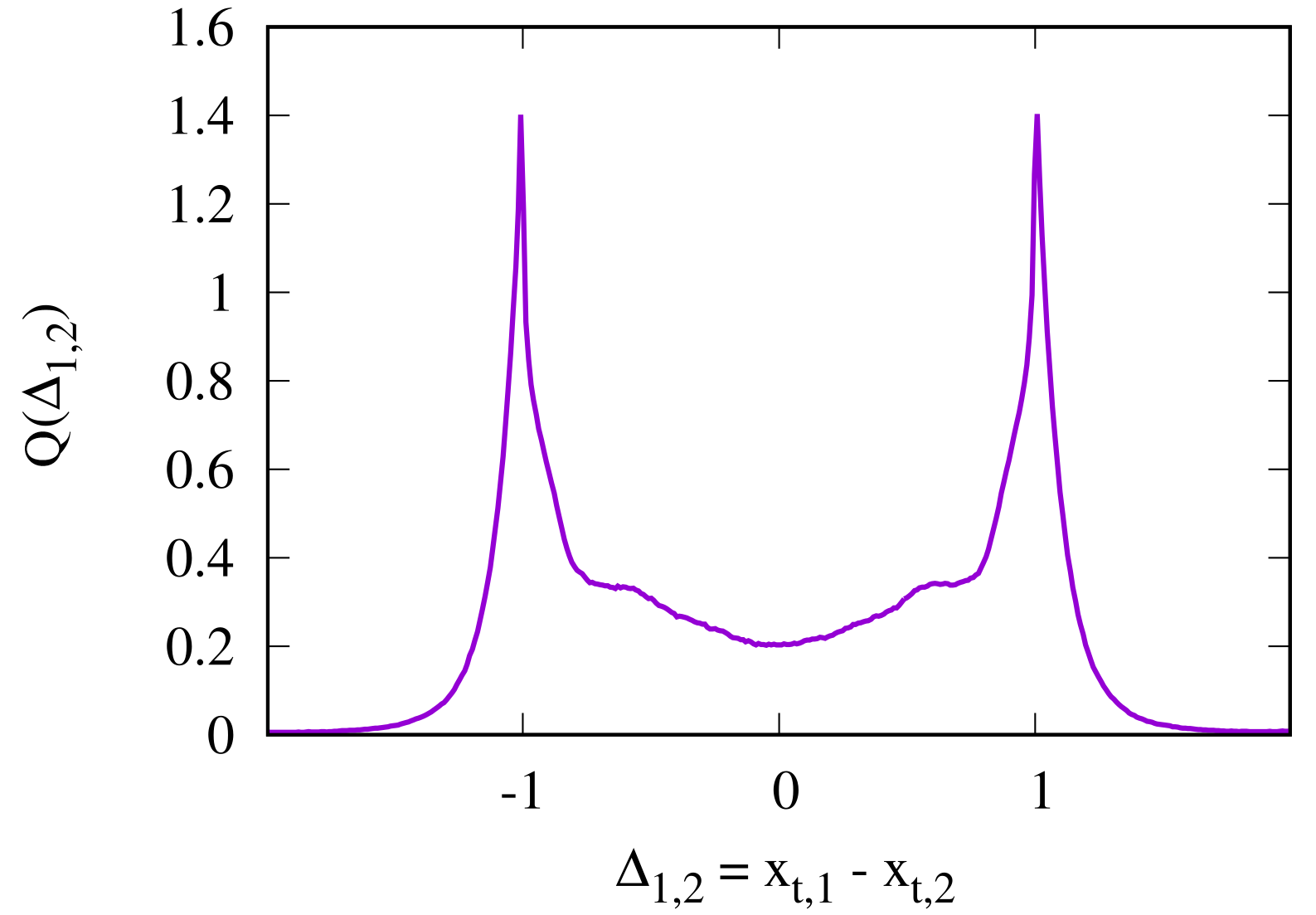}
\caption{Probability distribution of the tracer separation in the $x$ direction, $\D_{1,2} = x_{t,1} - x_{t,2}$, for two bound tracers in a narrow channel of length $L_x = 96.2$ and
width $L_y = 2.1 $. The average bath density is $\bar{\r} = 0.5$ and the force is $F = 16$.} 
\label{p_delta_channel}
\end{center}
\end{figure}

\begin{figure}
\begin{center}
\includegraphics[width=0.48\textwidth]{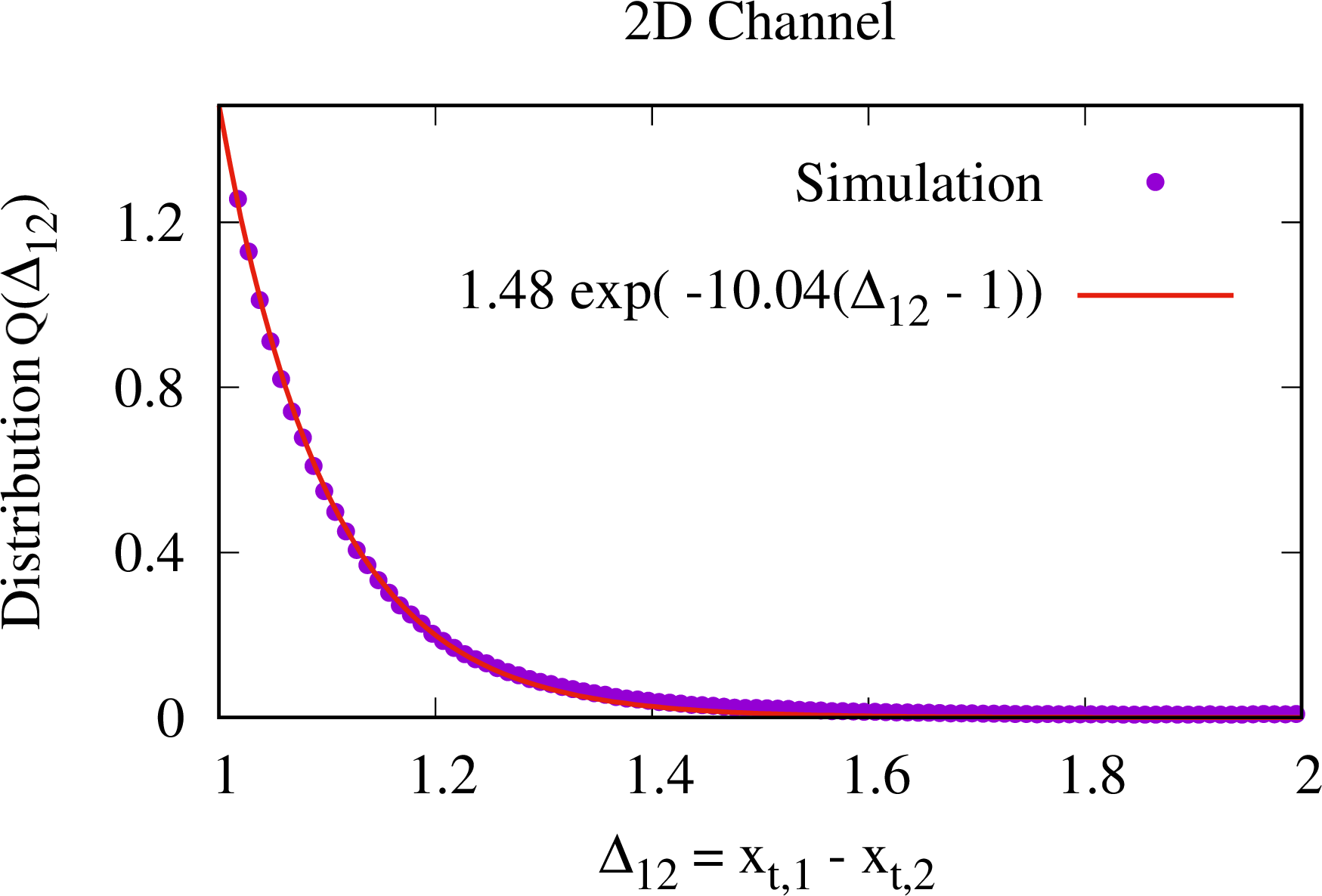}
\caption{A closeup of the region $ 1< \Delta_{1,2} < 2$ in
Fig. \ref{p_delta_channel} of the probability distribution   $Q(\Delta_{1,2})$. The violet dots show the simulated distribution and the red curve is a fit to exponential decay.}
\label{decay_1}
\end{center}
\end{figure}

Figure \ref{p_delta_channel} shows the stationary distribution  $Q(\D_{1,2})$  of the distance $\D_{1,2}$ between the two tracers.
The central domain of  $Q(\D_{1,2})$, i.e. for $-1< \D_{12} < 1$, corresponds to configurations where the two tracers are located one above the other, overlapping along the $x$ axis. Their distance distribution in this domain is determined by the specific details of their frequent mutual collisions. On the other hand, an exponential decay is clearly demonstrated for $|\D_{12}|>1$ in Fig. \ref{decay_1}.
This qualitatively agrees with the exponential behavior of the probability distribution $Q(k)$, obtained for the $1D$ lattice model in Eq. \eqref{eq:Q_k} and verified in Fig. \ref{fig:Q_k_fig}.


\begin{figure}
\begin{center}
\includegraphics[width=0.485\textwidth]{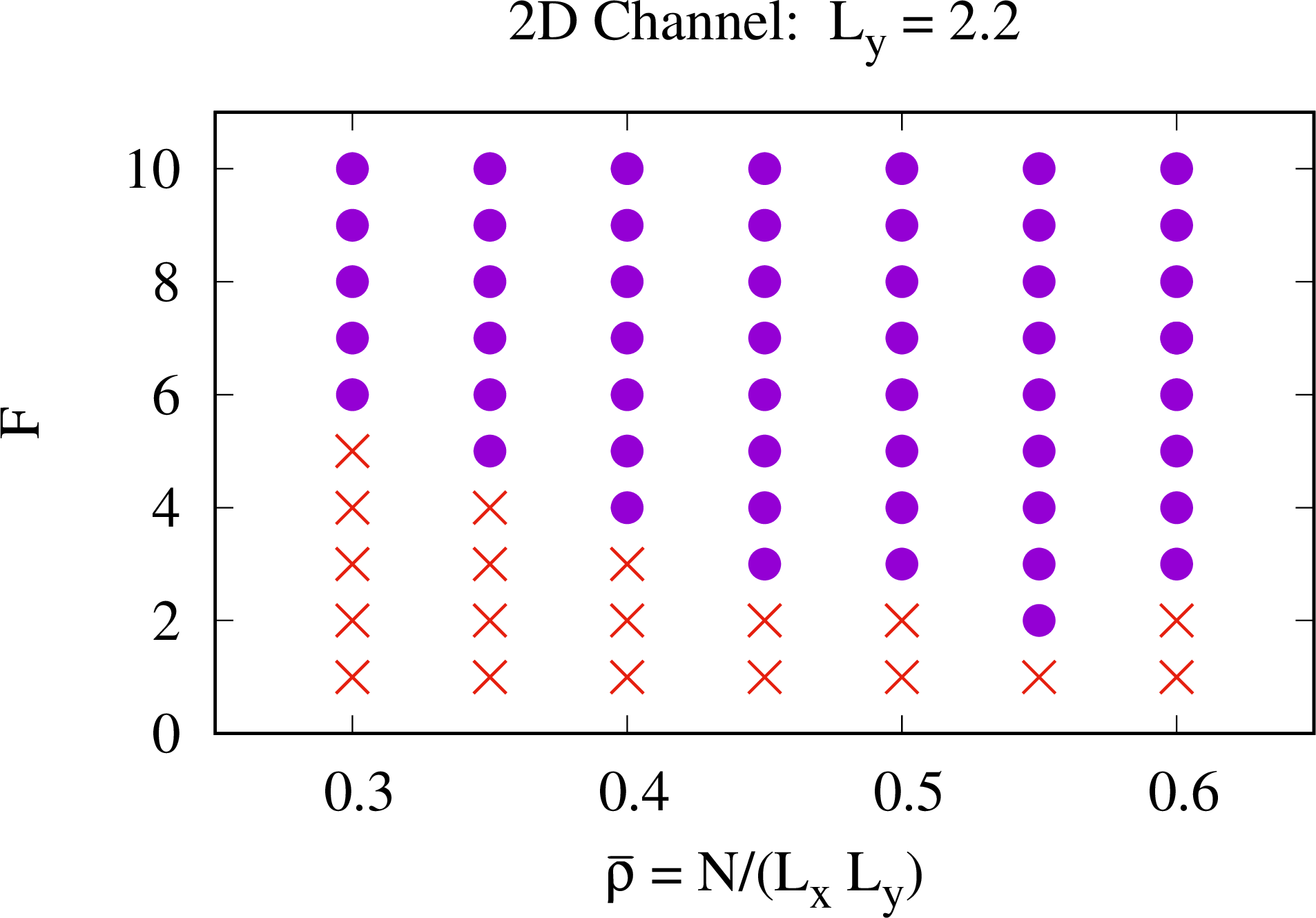}
\caption{Driving-force -- bath-density section of the phase diagram for a channel of width $L_{y}=2.2$ occupied by a total 
of $200$ particles, including the two tracers. Violet dots indicate parameter sets that yield attraction, whereas orange crosses indicate parameter sets for which there is no attraction.}
\label{parameter_section}
\end{center}
\end{figure}

Studying the 2D channel system with two tracers 
for a wide range of parameters, we obtain the $(\bar\r,F)$ phase diagram presented in Fig. \ref{parameter_section}. In this figure
attractive parameters are indicated by blue dots and repulsive parameters  by orange crosses. In an attractive state, generated by large-enough driving, the two tracers are observed to remain in close proximity to one another, following their initial encounter. If the driving force $F$ is reduced, the mean tracer separation gradually increases. When $F$ is lowered below a certain threshold, the two tracers fail to form a bound state, even after multiple encounters.


\section{Conclusions \label{sec:conclusions}}

Tracers driven along a crowded bath have been observed to display strong effective mutual attraction and condensation mediated by the bath particles. While this has been numerically demonstrated in a variety of models, a quantitative analysis of the attraction mechanism and the resulting effective potential, is not available.  
In the present paper we have analyzed the case of two driven tracers moving in a narrow channel, demonstrating that in the steady state, the tracers experience a strong confining potential. By introducing an effective $1D$ model, constructed to capture the dynamics of the tracers in the $2D$ channel, we show that the potential increases
linearly with the distance between the tracers, resulting in strongly bound tracers.


The model studied in this paper is that of a narrow $2D$ channel occupied by a thermal bath composed of $N$ Brownian disks with hard core interactions and two externally driven ''tracer'' disks. Although we explicitly consider a narrow channel, it is made to be wide enough to allow overtaking to occur. This implies that all particles, including the driven tracers, are able to move away from one another.

To analyze the attraction between the tracers, we modeled the $2D$ channel dynamics by that of a simple symmetric exclusion process (SSEP), extended to capture particle overtaking in the channel, on a $1D$ ring lattice. We have shown that, in the $1D$ lattice model, the dynamics of the two tracers can be regarded as that of two biased random walks whose moving rates are related to the stationary bath density profile (as seen in the tracers' reference frame). We found that in a region of the lattice model's dynamical parameters, the tracers  become effectively biased to move towards each other and form a robust bound state. Correspondingly, the distribution function of the two tracer's distance was shown to be exponentially decreasing with the distance, suggesting an effective linear attractive potential between the tracers. This effective interaction is mediated by the bath particles through their non-homogeneous steady state density profile. 
Extensive numerical simulations of the $2D$ narrow channel model were then shown to be in excellent qualitative agreement with these findings. In particular, the bath density profile, the tracer pairs' velocity, and the two tracers distance distribution were shown to exhibit the same properties analytically predicted for the $1D$ lattice model's localized phase.

\section{Acknowledgments \label{sec:Acknowledgments}}

We thank Bertrand Lacroix-A-Chez-Toine for helpful discussions. This work was supported by a research grant
from the Center of Scientific Excellence at the Weizmann Institute
of Science. The generous allocation of computer resources at the VSC3 cluster operated by Austrian Universities is gratefully acknowledged. 

\section*{Appendix - Bound State of Two Driven Tracers in the $1D$ Lattice Model \label{sec:appendix}}

This Appendix provides evidence which validate and support the claim that, when the two tracers form a bound pair in the model's localized phase, one may approximate them by an effective single-tracer that qualitatively behaves as predicted in Ref. \citep{Miron_2020}.

\begin{figure}
\includegraphics[scale=0.58]{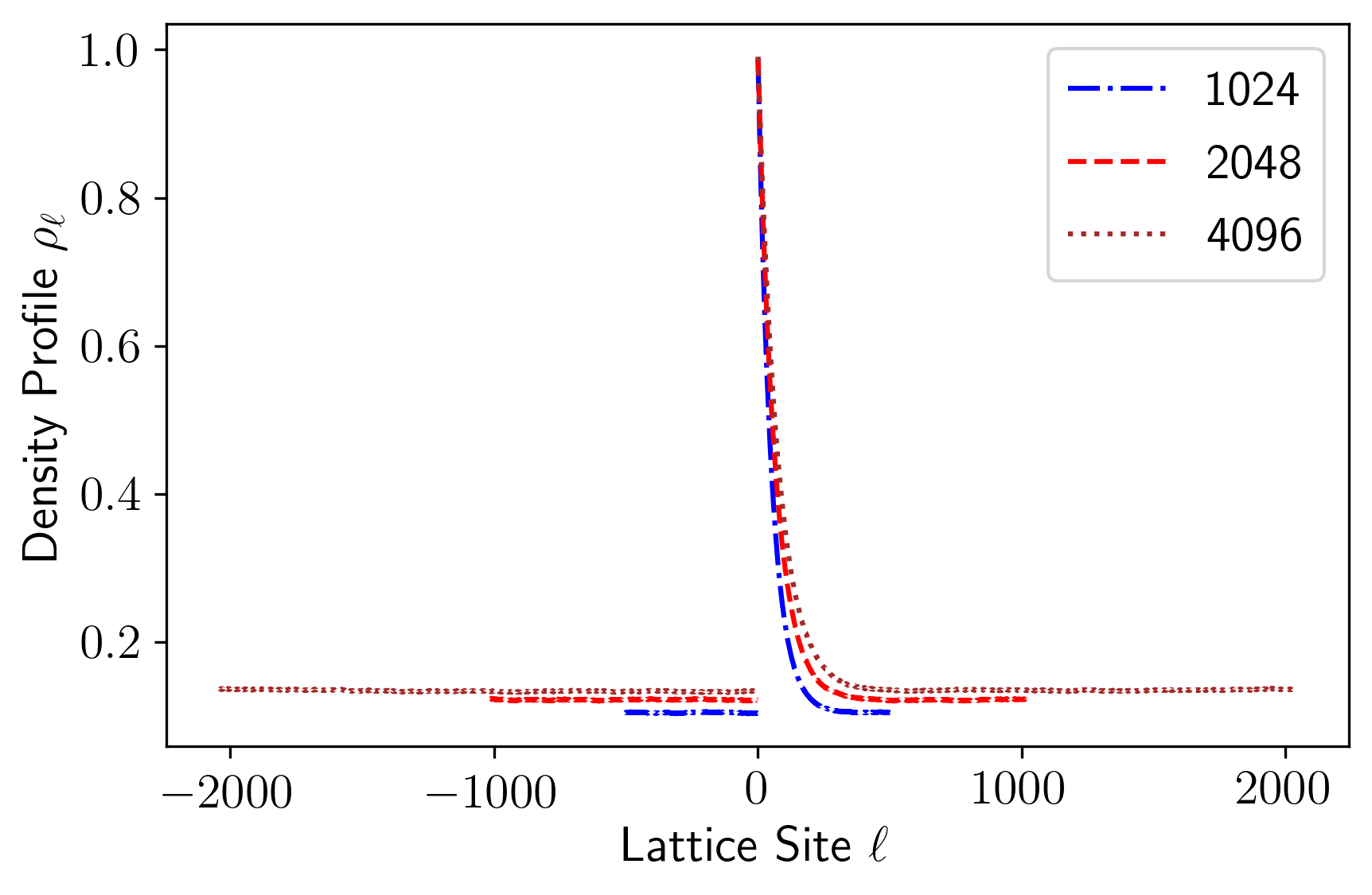}
\caption{Data collapse of the stationary bath density profile $\r_\el$ for two driven tracers and different system sizes $L$ in the $1D$ lattice model. The mean bath density is $\bar{\r} = 0.15$ and the dynamical rates are $p=1.9$, $q=0.1$, $p'=0.0075$, and $q'=0.0125$, corresponding to $\d=0.9$, $\d'=-0.25$, and $r/r'=100$.}
\label{fig:lattice_den}
\end{figure}

Figure \ref{fig:lattice_den1} shows the stationary bath density profile $\r_\el$ near the origin site $\el=0$ for both one and two driven tracers. In the case of two driven tracers, the right (i.e. positive) part of the profile is provided in the rightmost tracer's (i.e. tracer 1) reference frame while the left (i.e. negative) part is provided in the leftmost tracer's (i.e. tracer 2) frame. The figure shows close agreement of the density profiles near the tracers. A data collapse of the bath density profiles generated by two driven tracers is shown in Fig. \ref{fig:lattice_den} for different system sizes. As expected in the localized phase \citep{Miron_2020}, the deviation of the bath density profile from $\bar{\r}$ is localized around the origin and is independent of the system size $L$ as $L\ra\infty$.

\begin{figure}[H]
\includegraphics[scale=0.58]{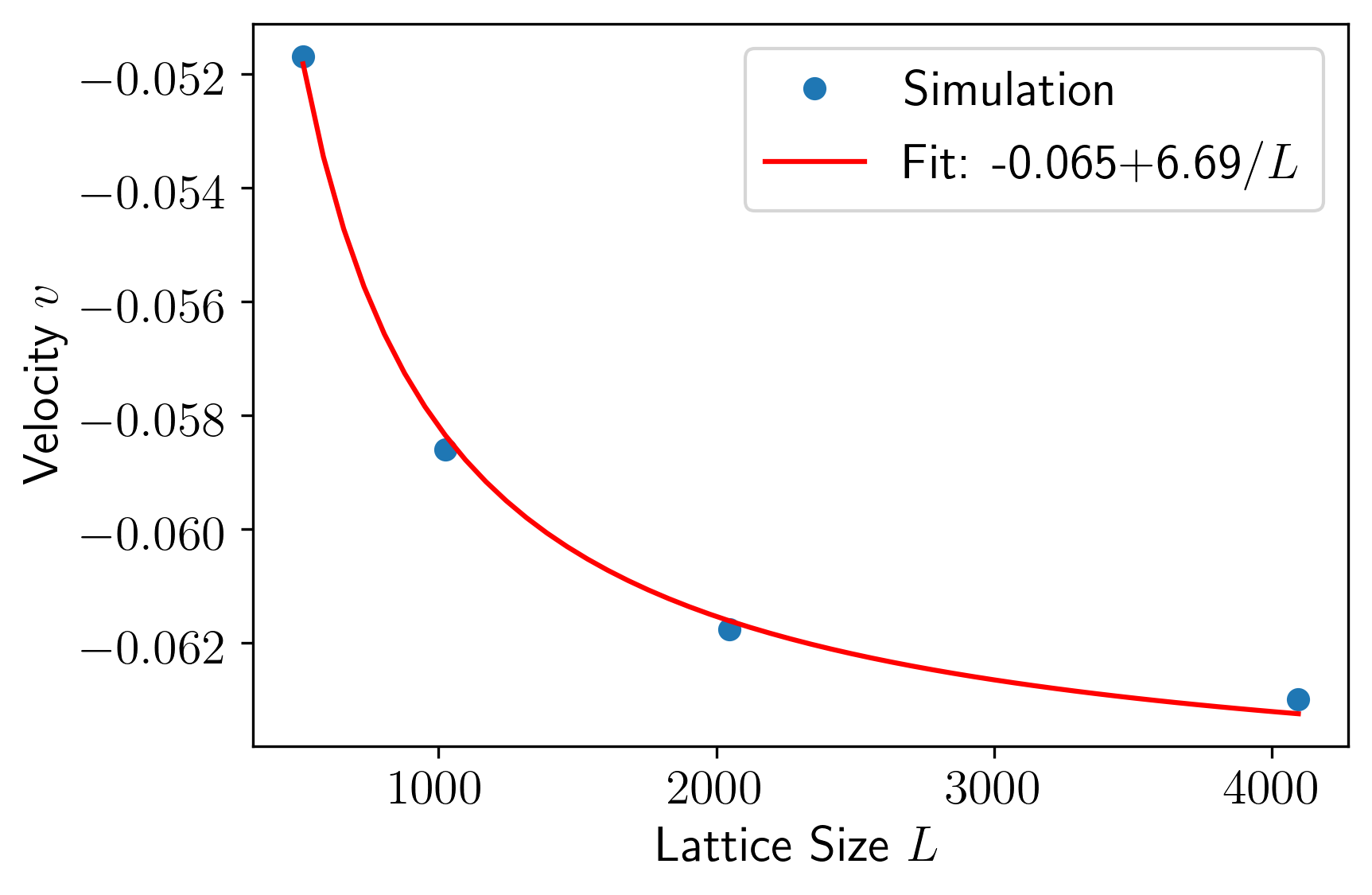}
\caption{The stationary velocity $v$ of the two-tracer bound pair versus system size $L$ in the $1D$ lattice model with a mean bath density of $\bar{\r}=0.15$ and dynamical rates $p=1.9$, $q=0.1$, $p'=0.0075$, and $q'=0.0125$, corresponding to $\d=0.9$, $\d'=-0.25$ and $r/r'=100$. Blue dots denote direct simulation results while the solid red curve is a fit to $c_{1}+c_{2}/L$.}
\label{fig:lattice_v}
\end{figure}

The velocity of a single driven tracer was also studied in \citep{Miron_2020}, where it was shown to attain a finite non-zero value as $L\ra\infty$. This appears to agree with Fig. \ref{fig:lattice_v}, which shows the stationary two-tracer bound pair velocity versus the system size $L$.

\end{document}